\documentclass[preprint,12pt]{elsarticle}

\usepackage{hyperref}
\usepackage{url}
\usepackage{graphicx}
\usepackage{amssymb}
\usepackage{amsmath}
\usepackage{units}
\usepackage[nooneline]{subfigure}
\usepackage{ptex2tex}
\biboptions{comma,super}

\newcommand{\refeq}[1]{(\ref{#1})}

\newcommand{\beq}{\begin{equation}}
\newcommand{\eeq}{\end{equation}}
\newcommand{\beqa}{\begin{eqnarray}}
\newcommand{\eeqa}{\end{eqnarray}}
\newcommand{\beqan}{\begin{eqnarray*}}
\newcommand{\eeqan}{\end{eqnarray*}}
\newcommand{\ben}{\begin{enumerate}}
\newcommand{\een}{\end{enumerate}}
\newcommand{\bit}{\begin{itemize}}
\newcommand{\eit}{\end{itemize}}
\newcommand{\ep}{\thinspace . }

\newcommand{\mathbfx}[1]{{\mbox{\boldmath $#1$}}}

\newcommand{\dif}{\, d}

\newcommand{\strainr}{\mathbfx{S}}

\newcommand{\x}{\mathbfx{x}}

\renewcommand{\u}{\mathbfx{u}}

\newcommand{\Id}{\mathbfx{I}}

\newcommand{\U}{\mathbfx{U}}
\newcommand{\V}{\mathbfx{V}}

\def\R{\mathbfx{R}}

\newcommand{\g}{\mathbfx{g}}

\newcommand{\f}{\mathbfx{f}}

\renewcommand{\v}{\mathbfx{v}}

\begin{document}
\begin{frontmatter}

\title{A FEniCS-Based Programming Framework for Modeling Turbulent Flow by the
Reynolds-Averaged Navier-Stokes Equations}

\author[ffi,simula]{Mikael Mortensen}
\ead{mikael.mortensen@ffi.no}

\author[simula,ifi]{Hans Petter Langtangen\corref{cor1}}
\ead{hpl@simula.no}
\cortext[cor1]{Corresponding author}

\author[cambridge]{Garth N. Wells}
\ead{gnw20@cam.ac.uk}

\address[ffi]{Norwegian Defence Research Establishment,
2007 Kjeller, Norway}

\address[simula]{Center for Biomedical Computing,
Simula Research Laboratory,
P.O.Box 134, 1325 Lysaker, Norway}

\address[ifi]{Department of Informatics, University of Oslo,
P.O.Box 1080 Blindern, 0316 Oslo, Norway}

\address[cambridge]{Department of Engineering, University of Cambridge,
Trumpington Street, Cambridge CB2 1PZ, United Kingdom}
\begin{abstract}
Finding an appropriate turbulence model for a given flow case usually
calls for extensive experimentation with both models and numerical
solution methods. This work presents the design and implementation
of a flexible, programmable software framework for assisting with
numerical experiments in computational turbulence.  The framework targets
Reynolds-averaged Navier-Stokes models, discretized by finite element
methods.  The novel implementation makes use of Python and the FEniCS
package, the combination of which leads to compact and reusable code,
where model- and solver-specific code resemble closely the mathematical
formulation of equations and algorithms.  The presented ideas and
programming techniques are also applicable to other fields that involve
systems of nonlinear partial differential equations.  We demonstrate
the framework in two applications and investigate the impact of various
linearizations on the convergence properties of nonlinear solvers for
a Reynolds-averaged Navier-Stokes model.
\end{abstract}
\begin{keyword}
Turbulent flow \sep
RANS models\sep
finite elements\sep
Python \sep
FEniCS \sep
object-oriented programming \sep
problem solving environment
\end{keyword}

\end{frontmatter}
\section{Introduction}

Turbulence is the rule rather than the exception when water flows in
nature, but finding the proper turbulence model for a given flow case
is demanding.  There exists a large number of different turbulence
models, and a researcher in computational turbulence would benefit
from being able to easily switch between models, combine models,
refine models and implement new ones. As the models consist of complex,
highly nonlinear systems of partial differential equations (PDEs),
coupled with the Navier-Stokes (NS) equations, constructing efficient
and robust iteration techniques is model- and problem-dependent, and
hence subject to extensive experimentation. Flexible software tools can
greatly assist the researcher experimenting with models and numerical
methods. This work demonstrates how flexible software can be designed
and implemented using modern programming tools and techniques.

Precise prediction of turbulent flows is still a very challenging task.
It is commonly accepted that solutions of the Navier-Stokes equations,
with sufficient resolution of all scales in space and time (Direct
Numerical Simulation, DNS), describe turbulent flow. Such an approach
is, nevertheless, computationally feasible only for low Reynolds
number flow and simple geometries, at least for the foreseeable
future.  Large Eddy Simulations (LES), which resolve large scale
motions and use subgrid models to represent the unresolved scales are
computationally less expensive than DNS, but are still too expensive
for the simulation of turbulent flows in many practical applications.
A computationally efficient approach to turbulent flows is to work with
Reynolds-averaged Navier-Stokes (RANS) models. RANS models involve solving
the incompressible NS equations in combination with a set of transport
equations for statistical turbulence quantities.  The uncertainty in
RANS models lies in the extra transport equations, and for a given flow
problem it is a challenge to pick an appropriate model. There is hence
a need for a researcher to experiment with different models to arrive
at firm conclusions on the physics of a problem.

Most commercial computational fluid dynamics (CFD) packages contain a
limited number of turbulence models, but allow users to add new models
through ``user subroutines'' which are called at each time level in
a simulation. The implementation of such routines can be difficult,
and new models might not fit easily within the constraints imposed
by the design of the package and the ``user subroutine'' interface.
The result is that a specific package may only support a fraction of
the models that a practitioner would wish to have access to.  There is
a need for CFD software with a flexible design so that new PDEs can be
added quickly and reliably, and so that solution approaches can easily
be composed. We believe that the most effective way of realizing such
features is to have a \emph{programmable} framework, where the models and
numerics are defined in terms of a compact, high-level computer language
with a syntax that is based on mathematical language and abstractions.

A software system for RANS modeling must provide higher-order spatial
discretizations, fine-grained control of linearizations, support
for both Picard and Newton type iteration methods, under-relaxation,
restart of models, combinations of models and the easy implementation
of new PDEs.  Standard building blocks needed in PDE software, such
as forming coefficient matrices and solving linear systems, can act
as black boxes for a researcher in computational turbulence.  To the
authors' knowledge, there is little software with the aforementioned
flexibility for incompressible CFD.  There are, however, many programmable
environments for solving PDEs. A non-exhaustive list includes
\citet{Cactus},
\citet{COMSOL},
\citet{www:deal.II} \citep{deal.II:paper},
\citet{Diffpack},
\citet{DUNE},
\citet{FEniCS}\citep{LoggWells2009a},
\citet{GetDP},
\citet{Getfempp},
\citet{OpenFOAM},
\citet{Overture},
\citet{PyADH}
and \citet{SAMRAI}.
Only a few of these packages have been extensively used for turbulent
flow.  \citet{OpenFOAM} is a well-structured and widely used
object-oriented C++ framework for CFD, based on finite volume methods,
where new models can quite easily be added through high-level C++
statements. \citet{Overture},\citep{Overture:ISCOPE97} is also an
object-oriented C++ library used for CFD problems, allowing complex
movements of overlapping grids.  \citet{PyADH} is a modern Python-
and finite element-based software environment for solving PDEs, and has
been used extensively for CFD problems, including free surface flow and
RANS modeling.  FEniCS \citep{FEniCS,FEniCS:book} is a recent C++/Python
framework, where systems of PDEs and corresponding discretization and
iteration strategies can be defined in terms of a few high-level Python
statements which inherit the mathematical structure of the problem,
and from which low level code is generated.  The approach advocated
in this work utilizes FEniCS tools.  All FEniCS components are freely
available under GNU general public licenses \citep{FEniCS}. A number
of application libraries that make use of the FEniCS software have
been published \citep{www:fenics-apps}. For instance, {\fontsize{11pt}{11pt}\texttt{cbc.solve}}
\citep{www:cbc.solve} is a framework for solving the incompressible
Navier-Stokes equations and the Rheology Application Engine (Rheagen)
\citep{www:rheagen} is a framework for simulating non-Newtonian flows. Both
applications share some of the features of the current work.

Traditional simulation software packages are usually implemented
in Fortran, C, or C++ because of the need for high computational
performance. A consequence is that these packages are less user-friendly
and flexible, but far more efficient, than similar projects implemented
in scripting languages such as Matlab or Python. In FEniCS, scripting
is combined with symbolic mathematics and code generation to provide
both user-friendliness and efficiency. Specifically, the Unified Form
Language (UFL), a domain-specific language for the specification of
variational formulations of PDEs, is embedded within the programming
language Python.  Variational formulations are then just-in-time compiled
into C++ code for efficiency. The generated C++ code can be expected
to outperform hand-written quadrature code since special-purpose PDE
compilers~\citep{SFC:10,KirbyLogg2008a,oelgaard:2010} are employed.
UFL has built-in support for automatic differentiation, derivation
of adjoint equations, etc., which makes it particularly useful for
complicated and coupled PDE problems.

Several authors have addressed how object-oriented and
generative programming can be used to create flexible
libraries for solving PDEs, but there are significantly fewer
contributions dealing with the design of frameworks on top of such
libraries for addressing multi-physics problems and coupling of PDEs
\citep{Peskin:Hardin:96,TCSE1,Larson:05,Joppich:06,Edwards:Sierra:04,Munthe:00,Rousin:patterns:10}.
These contributions focus on how the C++ or Fortran 90 languages can be
utilized to solve such classes of problems. This work builds on these
cited works, but applies Python as programming language and FEniCS as
tool for solving PDEs.  Python has strong support for dynamic classes
and object orientation, and since variables are not declared in Python,
generative programming comes without any extra syntax (in contrast with
templates in C++).  Presented code examples from the framework will
demonstrate how these features, in combination with FEniCS, result
in clean and compact code, where the specification of PDE models and
linearization strategies can be expressed in a mathematical syntax.

FEniCS supports finite element schemes, including discontinuous Galerkin
methods \citep{OelgaardLoggEtAl2008a}, but not finite difference methods.
Many finite volume methods can be constructed as low-order discontinuous
Galerkin methods using FEniCS \citep{wells:2010}.  Despite the development
of several successful methods for solving the NS equations and LES models
by finite element methods, finite element methods have not often been
applied to RANS models, though some research contributions exist in this
area \citep{Ilinca:1998,Benim:2005,Smith:2005,Lubon:09}.

The remainder of this paper is organized as follows.
Section~\ref{FEniCS} demonstrates the use of FEniCS for solving
simple PDEs and briefly elaborates some key aspects of FEniCS.
Section~\ref{math:models} presents a selection of PDEs which form
the basis of some common RANS models.  Finite element formulations
of a typical RANS model and the iteration strategies for handling
nonlinear equations appear in Section~\ref{numerics}. The
software framework for NS solvers and RANS models is described in
Section~\ref{sec:design}. Section~\ref{RANS:results} demonstrates
two applications of the framework and investigates the impact of
different types of linearizations. In
Section~\ref{sec:comp:efficiency} we briefly discuss the computational
efficiency of the framework, and some concluding perspectives are
drawn in Section~\ref{sec:conclusion}.  The code framework we describe,
cbc.rans, is open source and available under the Lesser GNU Public
license~\cite{www:cbc.rans}.

\section{FEniCS for solving differential equations}
\label{FEniCS}

FEniCS is a collection of software tools for the automated solution
of differential equations by finite element methods. FEniCS includes
tools for working with computational meshes, linear algebra and finite
element variational formulations of PDEs. In addition, FEniCS provides
a collection of ready-made solvers for a variety of partial differential
equations.
\subsection{Solving a partial differential equation}
\label{FEniCS:Poisson:numerics}

To illustrate how PDEs can be solved in FEniCS, we consider the
weighted Poisson equation $-\nabla\cdot (\kappa \nabla u) = f$ in some
domain $\Omega \subset \mathbb{R}^d$ with $\kappa = \kappa(x)$ a given
coefficient. On a subset of the boundary, denoted by $\partial\Omega_D$,
we prescribe a Dirichlet condition $u=0$, while on the remainder of the
boundary, denoted by $\partial\Omega_R$, we prescribe a Robin condition
$-\kappa{\partial u/\partial n}=\alpha (u-u_0)$, where $\alpha$ and $u_0$
are given constants.

To solve the above boundary-value problem, we first need to define the
corresponding variational problem. It reads: find $u\in V$ such that
\begin{equation}
  F\equiv \int_\Omega \kappa\nabla u\cdot \nabla
 v \dif x - \int_{\Omega}f v \dif x + \int_{\partial\Omega_R}
\alpha (u-u_0) v \dif s = 0 \quad\forall\, v\in V,
\label{Poisson:Feq0:cont}
\end{equation}
where $V$ is the standard Sobolev space $H^1(\Omega)$ with $u = v = 0$
on $\partial\Omega_D$.  The function $u$ is known as a trial function
and $v$ is known as a test function.  We can partition $F$ into a
``left-hand side'' $a(u,v)$ and a ``right-hand side'' $L(v)$,
\begin{equation}
   F = a(u,v) - L(v),
\end{equation}
where
\begin{align}
a(u,v) &=  \int_{\Omega} \kappa\nabla u \cdot \nabla v \dif x
         + \int_{\partial\Omega_R} \alpha u v \dif s,
\\
L(v)   &= \int_{\Omega} fv \dif x + \int_{\partial\Omega_R } \alpha u_0 v \dif s.
\end{align}

For numerical approximations, we work with a finite-dimensional subspace
$V_h \subset V$ and aim to find an approximation $u \in V_h$ such that
\begin{equation}
  a(u, v) = L(v) \quad \forall \ v\in V_h.
\label{Poisson:Feq0}
\end{equation}
This leads to a linear system $AU=b$, where $A_{ij} = a(\phi_j, \phi_i)$
and $b_i = L(\phi_i)$ are the matrix and vector obtained by evaluating
the bilinear form $a$ the and linear $L$ for the basis functions of the
discrete finite element function space and $U\in\mathbb{R}^N$ is the
vector of expansion coefficients for the finite element solution $u(x)
= \sum_{j=1}^N U_j \phi_j(x)$.

To solve equation \refeq{Poisson:Feq0} in FEniCS, all we have to
do is (i) define a mesh of triangles or tetrahedra over $\Omega$; (ii)
define the boundary segments $\partial\Omega_D$ and $\partial\Omega_R$
(only $\partial\Omega_D$ has to be defined in this case); (iii) define
the function space $V_h$; (iv) define $F$; (v) extract the left-hand
side $a$ and the right-hand side $L$; (vi) assemble the matrix $A$ and
the vector $b$ from $a$ and $L$, respectively; and (vii) solve the
linear system $AU=b$.
To be specific, we take
$d=2$,
$x=(x_0, x_1)$,
$\kappa(x_0,x_1) = x_1 \sin(\pi x_0)$,
$f(x_0,x_1)=0$,
$g(x_0,x_1)=0$,
and
$\alpha= 10$ and $u_0=2$.
The following Python program performs the above steps (i)--(vii):
\begin{Verbatim}[fontsize=\fontsize{11pt}{11pt},tabsize=8,baselinestretch=1.0]
from dolfin import *
mesh = Mesh('mydomain.xml.gz')
dOmega_D = MeshFunction('uint', mesh, 'myboundary.xml.gz')

V = FunctionSpace(mesh, 'Lagrange', degree=1)

g = Constant(0.0)
bc = DirichletBC(V, g, dOmega_D)

u = TrialFunction(V)
v = TestFunction(V)
f = Constant(0.0)
k = Expression('x[1]*sin(pi*x[0])')
alpha = 10; u0 = 2

F = inner(k*grad(u), grad(v))*dx - f*v*dx + alpha*(u-u0)*v*ds

a = lhs(F); A = assemble(a)
L = rhs(F); b = assemble(L)
bc.apply(A, b) # set Dirichlet conditions
solve(A, u.vector(), b, 'gmres', 'ilu')
plot(u)
\end{Verbatim}
\noindent
The FEniCS tools used in this program are imported from the {\fontsize{11pt}{11pt}\texttt{dolfin}}
package, which defines classes like {\fontsize{11pt}{11pt}\texttt{Mesh}}, {\fontsize{11pt}{11pt}\texttt{DirichletBC}},
{\fontsize{11pt}{11pt}\texttt{FunctionSpace}}, {\fontsize{11pt}{11pt}\texttt{TrialFunction}}, {\fontsize{11pt}{11pt}\texttt{TestFunction}}, and key
functions such as {\fontsize{11pt}{11pt}\texttt{assemble}}, {\fontsize{11pt}{11pt}\texttt{solve}} and {\fontsize{11pt}{11pt}\texttt{plot}}. We first
load a mesh and boundary indicators from files. Alternatively, the mesh
and boundary indicators can be defined as part of the program. The type of
discrete function space is defined in terms of a mesh, a class of finite
element (here {\fontsize{11pt}{11pt}\texttt{'Lagrange'}} means standard continuous Lagrange finite
elements~\citep{BrennerScott2008}) and a polynomial degree. The function
space {\fontsize{11pt}{11pt}\texttt{V}} used in the program corresponds to continuous piecewise
linear elements on triangles. In addition to continuous piecewise
polynomial function spaces, FEniCS supports a wide range of finite element
methods, including arbitrary order continuous and discontinuous Lagrange
elements, and arbitrary order $H(\rm div)$ and $H(\rm curl)$ elements.
The full range of supported elements is listed in \citet{LoggWells2009a}.

The variational problem is expressed in terms of the Unified Form Language
(UFL), which is another component of FEniCS. The key strength of UFL
is the close correspondence between the mathematical notation for~$F$
and its Python implementation~{\fontsize{11pt}{11pt}\texttt{F}}.  Constants and expressions can be
compactly defined and used as parts of variational forms. Terms multiplied
by {\fontsize{11pt}{11pt}\texttt{dx}} correspond to volume integrals, while multiplication
by {\fontsize{11pt}{11pt}\texttt{ds}} implies a boundary integral. Meshes may include several
subdomains and boundary segments, each with its corresponding volume
or boundary integral.  From the variational problem~{\fontsize{11pt}{11pt}\texttt{F}}, we may
use the operators {\fontsize{11pt}{11pt}\texttt{lhs}} and {\fontsize{11pt}{11pt}\texttt{rhs}} to extract the left- and
right-hand sides which may then be assembled into a matrix~{\fontsize{11pt}{11pt}\texttt{A}} and
a vector~{\fontsize{11pt}{11pt}\texttt{b}} by calls to the {\fontsize{11pt}{11pt}\texttt{assemble}} function. The Dirichlet
boundary conditions may then be enforced as part of the linear system
$AU = b$ by the call {\fontsize{11pt}{11pt}\texttt{bc.apply(A, b)}}. Finally, we solve the linear
system using the generalized minimal residual method ({\fontsize{11pt}{11pt}\texttt{'gmres'}})
with ILU preconditioning ({\fontsize{11pt}{11pt}\texttt{'ilu'}}).

\subsection{Solving a system of partial differential equations}
\label{FEniCS:Stokes:numerics}

The Stokes problem is now considered. It will provide a basis for
the incompressible Navier-Stokes equations in the following section.
The Stokes problem, allowing for spatially varying viscosity, involves
the system of equations
\begin{align}
- \nabla \cdot \nu (\nabla \u + \nabla \u^{T} ) + \nabla p &= \f,
\label{Stokes:eq1}
\\
\nabla\cdot\u &= 0.
\label{Stokes:eq2}
\end{align}
For the variational formulation, we introduce a vector test function
$\v \in \V$ for \refeq{Stokes:eq1} and a scalar test function $q \in
Q$ for \refeq{Stokes:eq2}. The trial functions are $\u \in \V$ and
$p \in Q$. Typically, $\V = [V]^d$, where $V$ is the space defined
for the Poisson problem, and $Q$ can be taken as the standard space
$L^2(\Omega)$. The corresponding variational formulation reads: find $(\u,
p)\in \V \times Q$ such that
\begin{multline}
  F \equiv  \int_{\Omega} \nu (\nabla\u + \nabla\u^{T}) : \nabla\v \dif x
 - \int_{\Omega} p \nabla\cdot\v \dif x
 - \int_{\Omega} \nabla \cdot\u \, q \dif x
\\
 - \int_{\Omega} \f \cdot \v \dif x
= 0 \quad \forall \ (\v,q) \in V \times Q.
\label{Stokes:eq:F}
\end{multline}
For simplicity, we consider in this example only problems where boundary
integrals vanish. As with the Poisson equation, we obtain a linear system
for the degrees of freedom of the discrete finite element solutions by
using finite-dimensional subspaces of $\V$ and~$Q$. Note the negative sign
in front of the third term in $F$. The sign of this term is arbitrary,
but it has been made negative such that the resulting matrix will be symmetric,
which is a feature that can be exploited by some preconditioners and
iterative solvers.

The following code snippet demonstrates the essential steps for solving
the Stokes problem in FEniCS:
\begin{Verbatim}[fontsize=\fontsize{11pt}{11pt},tabsize=8,baselinestretch=1.0]
V = VectorFunctionSpace(mesh, 'Lagrange', degree=2)
Q = FunctionSpace(mesh, 'Lagrange', degree=1)
VQ = V * Q # Taylor-Hood mixed finite element

u, p = TrialFunctions(VQ)
v, q = TestFunctions(VQ)
U = Function(VQ)

f = Constant((0.0, 0.0)); nu = Constant(1e-6)
F = nu*inner(grad(u) + grad(u).T, grad(v))*dx - p*div(v)*dx \
     - div(u)*q*dx - inner(f, v)*dx
A, b = assemble_system(lhs(F), rhs(F), bcs)
solve(A, U.vector(), b, 'gmres', 'amg_hypre')
u, p = U.split()
\end{Verbatim}
\noindent
We have for brevity omitted the code for loading a mesh, defining
boundaries and specifying Dirichlet conditions (the boundary conditions
are assumed to be available as a list {\fontsize{11pt}{11pt}\texttt{bcs}} in the program). A mixed
Taylor-Hood element is simply defined as {\fontsize{11pt}{11pt}\texttt{V*Q}}, where {\fontsize{11pt}{11pt}\texttt{V}} is a
second-order vector Lagrange element and {\fontsize{11pt}{11pt}\texttt{Q}} is a first-order scalar
Lagrange element.  Note that we solve for {\fontsize{11pt}{11pt}\texttt{U}}, which is a mixed
finite element function containing $\u$ and $p$. The function
{\fontsize{11pt}{11pt}\texttt{U}} can be split into individual finite element functions, {\fontsize{11pt}{11pt}\texttt{u}}
and {\fontsize{11pt}{11pt}\texttt{p}}, corresponding to $\u$ and~$p$.

More detailed information on the usage of and possibilities
with the FEniCS software suite can be found in the literature
\citep{KirbyLogg2006a,KirbyLogg2007a,Logg2007a,FEniCS:book,KirbyKnepleyEtAl2005a,LoggWells2009a,RognesKirbyEtAl2009a}.

\subsection{Automatic code generation}
\label{FEniCS:inside}

At the core of FEniCS is the C++/Python library
DOLFIN \citep{LoggWells2009a}, which provides data structures for finite
element meshes, functionality for I/O, a common interface to linear
algebra packages, finite element assembly, handling of parameters,
etc. DOLFIN differs from other finite element libraries in that it
relies on generated code for some core tasks. In particular, DOLFIN
relies on generated code for the assembly of finite element variational
forms. Code can be generated from a form expressed in UFL by one of
the two form compilers FFC \citep{KirbyLogg2006a} and SFC \citep{SFC:10}
that are available as part of FEniCS. The code may be generated prior
to compile-time by explicitly calling one of the form compilers, or
automatically at run-time (just-in-time compilation).  The latter is
the default behavior for users of the FEniCS Python interface.

\begin{figure}
  \center{\includegraphics[width=0.7\textwidth]{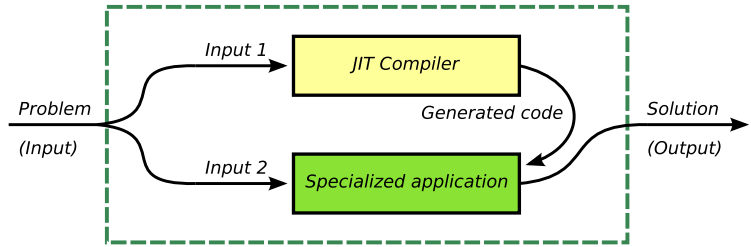}}
  \caption{Sketch of the code generation process
    in FEniCS. The user input is partitioned into two sets: data that
    requires special-purpose code such as finite element variational
    forms (Input~1), and data that can be handled efficiently by a
    general purpose routine such as the mesh, boundary conditions and
    coefficients (Input~2). For the first set of data, FEniCS calls a
    just-in-time (JIT) compiler to generate special-purpose code that
    may then be executed for the remaining set of data (Input~2) to
    compute the solution.}
\label{fig:codegen}
\end{figure}

Relying on generated code means that FEniCS is able to satisfy two
seemingly contradictory objectives: generality, by being capable of
generating code for a large class of linear and nonlinear finite
element variational problems, and efficiency by calling highly
optimized code generated for each specific variational problem given
by the user as input. This is illustrated in Figure~\ref{fig:codegen}.
It has been demonstrated
\citep{KirbyLoggEtAl2006a,KirbyKnepleyEtAl2005a,KirbyLogg2006a,KirbyLogg2007a,KirbyLogg2008a,oelgaard:2010}
that using form compilers permits the application of optimizations
and representations that could not be expected in handwritten code.

\section{Reynolds averaged Navier-Stokes equations}
\label{math:models}

Turbulent flows are described by the NS equations.
On a domain $\Omega \subset \mathbb{R}^d$
for time $t \in (0, t_s]$, the incompressible NS equations read
\begin{align} \label{eq:NS}
     \frac{\partial \U}{\partial t} + \U \cdot \nabla\U
      &= - \frac{1} {\varrho}\nabla P   + \nabla \cdot \nu(\nabla \U + \nabla \U^{T})  + \f,
\\
 \label{eq:cont}
 \nabla \cdot \U &= 0,
\end{align}
where $\U(\x,t)$ is the velocity, $P(\x,t)$ is the pressure, $\nu$ is the
kinematic viscosity, $\varrho$ is the mass density and $\f$ represents
body forces.  The incompressible NS equations must be complemented by
initial and appropriate boundary conditions to complete the problem.

Simulations of turbulent flows are usually computationally expensive
because of the need for extreme resolution in both space and
time. However, in most applications the average quantities are
of interest.  In the statistical modeling of turbulent flows, the
velocity and pressure are viewed as random space-time fields which
can be decomposed into mean and fluctuating parts: $\U = \u + \u'$
and $P = p + p'$, where $\u$ and $p$ are ensemble averages of $\U$ and
$P$, respectively, and $\u'$ and $p'$ are the random fluctuations about
the mean field.  Inserting these decompositions into \refeq{eq:NS} and
\refeq{eq:cont} results in a system of equations for the mean quantities
$\u$ and~$p$:
\begin{align}
 \label{eq:NS:m}
     \frac{\partial \u} {\partial t} + \u \cdot \nabla\u
      &= - \frac{1} {\varrho}\nabla p + \nabla \cdot \nu( \nabla \u + \nabla \u^{T})
            - \nabla \cdot \overline{\u' \otimes \u'}  + \f,
\\
 \label{eq:cont:m}
 \nabla \cdot \u &= 0,
\end{align}
where $\R = \overline{\u' \otimes \u'}$, known as the Reynolds stress
tensor, is the ensemble average of $\u' \otimes \u'$.  The Reynolds
stress tensor is unknown and solving equations \refeq{eq:NS:m}
and \refeq{eq:cont:m} requires approximating $\R$ in terms of $\u$,
$\nabla\u$, or other computable quantities.

A general observation on turbulence is that it is dissipative. This
observation has led to the idea of relating the Reynolds stress
to the strain rate tensor of the mean velocity field,
$\strainr = (\nabla\u + \nabla\u^T)/2$.  More specifically,
\beq
\R = -2\nu_T\strainr + {\frac{2}{3}}k\Id,
\eeq
where $\nu_T$ is the ``turbulent viscosity'', $k =
\overline{\u' \cdot \u'}/2$ is the turbulent kinetic energy
and $\Id$ is the identity tensor.  Many models have been proposed for
the turbulent viscosity.  The most commonly employed ``one-equation''
turbulence model is that described by \citet{Spalart:92}.  It involves a
transport equation for a ``viscosity'' parameter, coupled to 11 derived
quantities with 9 model parameters.

Two-equation turbulence models represent the largest class of RANS
models, providing two transport equations for the turbulence length
and time scales. This family of models includes the $k$--$\epsilon$
models \citep{jones1972,launder1974} and the $k$--$\omega$ models
\citep{Wilcox:1988}.  Of the two-equation models, we limit our
considerations in this work to $k$--$\epsilon$ models. Due to severe
mesh resolution requirements these models usually involve the use
of wall functions instead of regular boundary conditions on solid
walls. Support for the use of wall functions has been implemented in
the current framework and special near-wall modifications are employed
both for the standard $k$--$\epsilon$ model and the more elaborate four
equation $v^2$--$f$ model \citep{book:durbin}. Implementation aspects of
these wall modifications, though, involves a level of detail that is beyond
the scope of the current presentation. For this reason we will mainly focus on
a particular family of $k$--$\epsilon$ models, the low-Reynolds models,
that apply standard Dirichlet boundary conditions on solid walls.

The ``pseudo'' rate of dissipation of turbulent kinetic energy is defined
as\citep{Pope00}
\beq
\epsilon = \nu \overline{\nabla \u' : \nabla \u'}.
\label{epsilon:def}
\eeq
All $k$--$\epsilon$ models express the turbulent viscosity parameter
$\nu_T$ in terms of $k$ and $\epsilon$ (from dimensional arguments,
$\nu_T \sim k^2/\epsilon$).  The fluctuations $\u'$ are unknown,
and consequently $k$ and $\epsilon$ must be modeled.  A low-Reynolds
$k$--$\epsilon$ model in general form reads
\begin{align}
 \label{eq:RANS:m}
     \frac{\partial \u} {\partial t} + \u \cdot \nabla\u
      &= \nabla \cdot \nu_u (\nabla \u + \nabla \u^{T}) - \frac{1} {\varrho}\nabla p + \f,
\\
 \label{eq:cont2:m}
 \nabla \cdot \u &= 0,
\\
{\frac{\partial k}{\partial t}}
+ \u\cdot\nabla k &= \nabla \cdot (\nu_k  \nabla k) +P_k-\epsilon-D,
\label{k:eq:LS}
\\
{\frac{\partial\epsilon}{\partial t}} + \u\cdot\nabla\epsilon
    &= \nabla \cdot ( \nu_\epsilon \nabla \epsilon )
      + \left( C_{\epsilon 1} P_k - C_{\epsilon 2} f_2 \epsilon \right) \frac{\epsilon}{k}+E,
\label{eps:eq:LS}
\\
\nu_u &= \nu + \nu_T,
\label{nuu:eq:LS}
\\
\nu_k &= \nu + \frac{\nu_T}{\sigma_k},
\label{nuk:eq:LS}
\\
\nu_\epsilon &= \nu + \frac{\nu_T}{\sigma_\epsilon},
\label{nueps:eq:LS}
\\
\nu_T &= C_\mu f_{\mu} {\frac{k^2}{\epsilon}},
\label{nuT:eq:LS}
\\
P_k &= \R:\nabla\u,
\label{Pk:eq:LS}
\end{align}
where various terms which are model-specific are defined in
Table~\ref{tab:lowre} for three common low-Reynolds models.
The pressure $p$ in the NS equations
is a modified pressure that includes the kinetic energy from
the model for the Reynolds stresses. The dissipation rate term
$\epsilon$ is also modified and the pseudo-dissipation rate defined
in \eqref{epsilon:def} can be recovered from these models
as $\epsilon + D$. This modification is introduced in all low-Reynolds
models to allow a homogeneous
Dirichelt boundary condition for $\epsilon$ on solid walls.
Further discussion
of boundary conditions for RANS models are delayed until the
presentation of examples in Section~\ref{RANS:results}.
\begin{table}
\begin{tabular}{c|ccc}
           & \citet{chien1982} & \citet{launder1974} & \citet{jones1972} \\
\hline
$C_{\mu}$  & 0.09  & 0.09           & 0.09 \\
$\sigma_k$ & 1     & 1              & 1 \\
$\sigma_{\epsilon}$ & 1.3     & 1.3       & 1.3 \\
$D$          & $2\nu \frac{k}{y^2}$ & $2\nu | \nabla \sqrt{k}|^2$ & $2\nu | \nabla \sqrt{k}|^2$ \\
$E$          & $-\frac{2\nu \epsilon}{y^2} \exp \left( -0.5 y^+ \right)$ & $2\nu \nu_T |\nabla^2 \u|^2$ & $2\nu \nu_T |\nabla^2 \u|^2$ \\
$C_{\epsilon 1}$ & 1.35 & 1.44 & 1.55 \\
$C_{\epsilon 2}$ & 1.8 & 1.92 & 2.0 \\
$f_{\mu}$ & $1-\exp(-0.0115 y^+)$ & $\exp \left( \frac{-3.4}{(1+Re_T/50)^2} \right)$ & $\exp \left( \frac{-2.5}{(1+Re_T/50)} \right)$ \\
$f_2$ & $1-0.22 \exp\left(-\frac{Re_T^2}{36}\right)$ & $1-0.3\exp\left(-{Re}_T^2\right)$ & $1-0.3\exp\left(-{Re}_T^2\right)$
\end{tabular}
\caption{Various model constants and damping functions for three low-Reynolds
number turbulence models. $Re_T=k^2/(\nu \epsilon)$.}
\label{tab:lowre}
\end{table}

In the original models of \citet{jones1972} and \citet{launder1974}
$D =  2\nu({\partial \sqrt{k}}/{\partial y})^2$
and
$E = 2\nu \nu_T( {\partial^2 u_x}/{\partial y^2} )$,
where $y$ is the wall normal direction and $u_x$ is the mean velocity
tangential to the wall. To eliminate the coordinate dependency, we have
generalized these terms to the ones seen in Table~\ref{tab:lowre}.
The rationale behind the generalization is that $D$ and $E$ are only
important in the vicinity of walls, where the term $\partial u_{x} /
\partial y$ will be dominant.  The generalized terms will therefore
approach the terms in the original model in the regions where the terms
are significant, regardless of the geometry of the wall.
\section{Numerical methods for the Reynolds averaged Navier-Stokes
equations and models}
\label{numerics}

This section addresses numerical solution methods for the Navier-Stokes
equation presented in \eqref{eq:RANS:m} and~\eqref{eq:cont2:m} and the
turbulence models presented in Section~\ref{math:models}.  RANS models
are normally considered in a stationary setting, i.e., the mean flow
quantities do not depend on time. Hence, we will here ignore the time
derivatives appearing in the equations in Section~\ref{math:models},
even though we have also implemented solvers for transient systems
within the current framework. We also adopt the strategy of splitting
the total system of PDEs into (i) the Navier-Stokes system for $\u$
and $p$, with $\R$ given; and (ii) a system of equations for $\R$,
with $\u$ and $p$ given.
\subsection{Navier-Stokes solvers}
\label{NS:numerics}

There are numerous approaches to solving the NS equations. A
common choice is a projection or pressure correction scheme
\citep{NSreview:02,Donea:Huerta:2003}.  Here we present a
solver in which $\u$ and $p$ are solved in a coupled fashion.
The variational form consists of that for the Stokes problem in
Section~\ref{FEniCS:Stokes:numerics}, with an additional momentum
convection term. With $\varrho$ absorbed into $p$, the variational problem
for the NS equations reads: find $(\u, p) \in \V \times Q$ such that
\begin{multline}
F \equiv
       \int_{\Omega} (\u\cdot\nabla\u) \cdot  \v \dif x
     + \int_{\Omega} \nu_{u}(\nabla \u + \nabla \u^{T}) : \nabla\v \dif x
     - \int_{\Omega} p \nabla\cdot\v \dif x
\\
     - \int_{\Omega} (\nabla\cdot\u) q \dif x
     - \int_{\Omega} \f \cdot \v \dif x
= 0 \quad \forall \ (\v, q) \in \V \times Q.
\label{NS:varform:coupled}
\end{multline}
For low ``cell'' Reynolds numbers, equation~\refeq{NS:varform:coupled}
is stable provided appropriate finite element bases are used for $\u$ and
$p$.  For example, the Taylor-Hood element, with continuous second-order
Lagrange functions for the velocity and continuous first-order Lagrange
functions for the pressure is stable.  It may sometimes be advantageous
to use equal-order basis functions for the velocity and pressure field,
in which case a stabilizing term must be added to the equations to
control spurious pressure oscillations. Consider the residual of the
Navier-Stokes momentum equation~\eqref{eq:RANS:m}:
\begin{equation}
  \mathcal{R} \equiv \u \cdot \nabla \u + \nabla p
     - \nabla \cdot \nu_u (\nabla \u + \nabla \u^{T}) - \f.
\label{eq:residual}
\end{equation}
We choose to add the momentum residual, weighted by $\nabla q$, to the
variational formulation in~\eqref{NS:varform:coupled}, which yields the
pressure-stabilized problem: find $(\u, p) \in \V \times Q$ such that
\begin{multline}
F_{\rm stab} \equiv
       \int_{\Omega} (\u\cdot\nabla\u) \cdot  \v \dif x
     + \int_{\Omega} \nu_{u}(\nabla \u + \nabla \u^{T} ): \nabla\v \dif x
     - \int_{\Omega} p \nabla\cdot\v \dif x
\\
     - \int_{\Omega} (\nabla\cdot\u) q \dif x
     +  \int_{\Omega} \tau \mathcal{R}(\u, p) \cdot \nabla q  \dif x
     - \int_{\Omega} \f \cdot \v \dif x
\\
= 0 \quad \forall \ (\v, q) \in \V \times Q,
\label{eq:NS:varform:coupled_stabilized}
\end{multline}
where $\tau$ is a stabilization parameter, which is usually taken
to be $\beta h^{2}/4\nu$, where $\beta$ is a dimensionless parameter
and $h$ is a measure of the finite element cell size.  This method of
stabilizing incompressible problems is known as a pressure-stabilized
Petrov-Galerkin method (see \citet{Donea:Huerta:2003} for background).
The stabilizing
terms are residual-based, i.e., the stabilizing term vanishes for the
exact solution, hence consistency of the formulation is not violated.
Additional stabilizing terms would be required to avoid spurious velocity
oscillations in the direction of the flow if the cell-wise Reynolds
number is large.

Since the convection term (the first term in $F$) is nonlinear, iterations
over linearized problems are required to solve this problem. The
simplest linearization is a Picard-type method, also known as successive
substitution, where a previously computed solution $\u_{-}$ is used
for the advective velocity, i.e., $\int_{\Omega} (\u \cdot \nabla\u)
\cdot \v \dif x$ becomes $\int_{\Omega} (\u_{-} \cdot \nabla\u) \cdot
\v \dif x$ in the linearized problem,
\begin{multline}
\tilde F \equiv
   \int_{\Omega} (\u_{-}\cdot\nabla\u) \cdot  \v \dif x
 + \int_{\Omega} {\nu_u}_{-}(\nabla \u + \nabla \u^{T}) : \nabla\v \dif x
 - \int_{\Omega} p \nabla\cdot\v \dif x
\\
 - \int_{\Omega} (\nabla\cdot\u) q \dif x
 - \int_{\Omega} \f \cdot \v \dif x.
\label{NS:varform:coupled:picard}
\end{multline}
The linear system arising from setting $\tilde F=0$ is solved for a
new solution $\x_*$ but this solution is only taken as a tentative
quantity. Relaxation with a parameter $\omega$ is used to compute the
new approximation:
\beq
\x_{-} \leftarrow (1-\omega)\x_{-} + \omega \x_{*},
\label{linsol:Picard:relax}
\eeq
where $\x_{-} = (\u_{-}, p_{-})$. Under-relaxation with $\omega < 1$
may be necessary to obtain a convergent procedure.

Faster, but possibly less robust convergence can be obtained by employing
a full Newton method, which requires differentiation of $F$ with respect
to $\u$ to form the Jacobian $J$. Since $J$ and $F$ contain the most
recent approximations to $\u$ and $p$, we add the subscript ``$-$''
($J_{-}$, $F_{-}$).  In each iteration, the linear system $J_{-}\delta\x =
-F_{-}$ must be solved.  The correction $\delta\x$ is added to $\x_{-}$,
with a relaxation factor $\omega$, to form a new solution:
\beq
\x_{-} \leftarrow \x_{-} - \omega\delta\x,
\eeq
where again $\x_{-} = (\u_{-}, p_{-})$ and $\delta\x = (\delta\u, \delta
p)$.  Once $\u$ and $p$ have been computed, derived quantities, such as
 $\nabla\cdot\u$ and $\strainr$, can be evaluated.
\subsection{Turbulence models}
\label{sec:G:term}

The equations for $k$ and $\epsilon$, as presented in
Section~\ref{math:models}, need to be cast in a weak form for
finite element analysis.  The weak equations for \eqref{k:eq:LS} and
\eqref{eps:eq:LS} read: find $k\in V_k$ such that
\begin{multline}
F_k \equiv
 - \int_{\Omega} \u\cdot\nabla k v_k \dif x
 - \int_{\Omega} \nu_k\nabla k  \cdot \nabla v_k \dif x
 + \int_{\Omega} P_k v_k \dif x
\\
 - \int_{\Omega} \epsilon v_k \dif x
 -  \int_{\Omega} D v_{k} \dif x
 = 0 \quad \forall \ v_k\in V_k,
\label{k:eq:LS:dis}
\end{multline}
and find $\epsilon\in V_\epsilon$ such that
\begin{multline}
F_\epsilon \equiv
  - \int_{\Omega} \u\cdot\nabla\epsilon v_\epsilon \dif x
 - \int_{\Omega} \nu_\epsilon\nabla\epsilon \cdot \nabla v_\epsilon \dif x
\\
 + \int_{\Omega} \left( C_{\epsilon 1}P_k - f_2 C_{\epsilon 2}
         \epsilon \right) \frac{\epsilon}{k} v_\epsilon \dif x
+ \int_{\Omega} E v_\epsilon \dif x
=0 \quad \forall\, v_\epsilon\in V_{\epsilon},
\label{eps:eq:LS:dis}
\end{multline}
where $V_{k}$ and $V_{\epsilon}$ are suitably defined function spaces.
A natural choice is to set $V_k=V_\epsilon=V$, where $V$ is the space
suitable for the Poisson equation.  The above weak forms correspond
to $\partial k / \partial n = 0$ or prescribed $k$ on the boundary,
and $\partial \nu_{\epsilon} \epsilon / \partial n = 0$ or prescribed
$\epsilon$ on the boundary. As stated earlier, precise boundary
conditions will be defined in Section~\ref{RANS:results}.

For the Jones-Launder and Launder-Sharma models, the term $E$ requires
some special attention in a finite element context.  The term $E$
is proportional to~$|\nabla^{2} \u|^{2}$.  To avoid the difficulties
associated with the presence of second-order spatial derivatives when
using a finite element basis that possesses only $C^{0}$ continuity,
we introduce an auxiliary vector field $\g$, and project $\nabla^2\u$
onto it. The $\g$ field is computed by a finite element formulation for
$\g = \nabla^2 \u$, which reads: find $\g \in \V$ such that
\beq
 \int_{\Omega} \g \cdot \v \dif x
   = - \int_{\Omega} \nabla\u : \nabla\v \dif x
     + \int_{\partial \Omega} {\frac{\partial\u}{\partial n}} \cdot  \v \dif s
  \quad \forall \v \in \V.
\label{auxL:eq}
\eeq
Then, $E$ in~\eqref{eps:eq:LS:dis} can be computed using $\g$ rather
than~$u$ directly. It should be pointed out that the need to do a variety
of standard and special-purpose projections can arise frequently when solving
multi-physics problems. The ease with which we can perform this operation is
perhaps one of the less obvious attractive features of having a framework
built around abstract variational formulations.
\subsubsection{Segregated and coupled solution approaches}

The equations for $k$ and $\epsilon$ are usually solved in sequence,
which is known as a segregated approach.  The first problem involves:
given $\u \in \V$ and $\epsilon \in V_\epsilon$, find $k\in V_k$ such that
\begin{equation}
F_k = 0 \quad \forall \, v_k\in V_k,
\end{equation}
and then given $\u \in \V$ and $k \in V_{k}$, find $\epsilon\in
V_\epsilon$ such that
\begin{equation}
F_\epsilon = 0 \quad \forall\, v_\epsilon\in V_{\epsilon}\ep
\end{equation}
Alternatively, the two equations of the $k$--$\epsilon$ system can be
solved simultaneously, which we will refer to as a coupled approach.
The variational statement reads: find $(k,\epsilon)\in V_k\times
V_\epsilon$ such that
\beq
F_k + F_\epsilon = 0\quad\forall\, (v,q)\in V_k \times V_{\epsilon}.
\label{k:eps:coupled:dis}
\eeq
\subsubsection{Solving the nonlinear equations}

Nonlinear algebraic equations arising from nonlinear variational forms are
solved by defining a sequence of linear problems whose solutions hopefully
converge to the solution of the underlying nonlinear problem. Let the
subscript ``$-$'' indicate the evaluation of a function at the previous
iteration, e.g.  $s_{-}$ is the value of $s$ in the previous iteration,
and let $s$ be the unknown value in a linear problem to be solved at the
current iteration.  For derived quantities, like $\nu_{k-}$ and $P_{k-}$,
the subscript indicates that values at the previous iteration are used
in evaluating the expression.

\paragraph{Picard iteration}
We regard \refeq{k:eq:LS:dis} as an equation for $k$ and use the previous
iteration value $\epsilon_{-}$ for $\epsilon$.  Other nonlinearities can
be linearized as follows (the tilde in $\tilde F_k$ denotes a linearized
version of $F_k$):
\begin{multline}
\tilde F_k \equiv
 - \int_{\Omega} \u\cdot\nabla k  \, v_k \dif x
 - \int_{\Omega} \nu_{k-} \nabla k \cdot \nabla v_k \dif x
 + \int_{\Omega} P_{k-} v_k \dif x
\\
 - \int_{\Omega} \epsilon_{-} \frac{k}{k_{-}} v_k \dif x
 - \int_{\Omega} D_{-} v_k \dif x,
\label{k:eq:LS:dis:lin}
\end{multline}
where for the Launder--Sharma and Jones--Launder models
\begin{equation}
D_{-} =  {\frac{1}{2}}\nu|k_{-}|^{-1} \nabla k_{-}\cdot\nabla k_{-}\ep
\end{equation}
Note the introduction of $k/k_{-}$ in the term involving~$\epsilon$.  This
is to allow for the implicit treatment of this term .  The corresponding
linear version of \refeq{eps:eq:LS:dis} reads
\begin{multline}
\tilde F_\epsilon \equiv
-\int_{\Omega} \u\cdot\nabla\epsilon \, v_\epsilon \dif x
- \int_{\Omega} \nu_{\epsilon-} \nabla\epsilon \cdot \nabla v_\epsilon \dif x
\\
 + \int_{\Omega} \left( C_{\epsilon 1}P_{k-} - f_{2-} C_{\epsilon 2}
              \epsilon \right) \frac{\epsilon_{-}}{k_{-}} v_\epsilon \dif x
+ \int_{\Omega} E_{-} v_\epsilon \dif x,
\label{eps:eq:LS:dis:lin}
\end{multline}
where
\begin{equation}
  E_{-} =  2\nu \nu_{T-} |\g |^2\ep
\end{equation}
When solving \refeq{eps:eq:LS:dis:lin}, we have the possibility of
using the recently computed $k$ value from \refeq{k:eq:LS:dis:lin} in
expressions involving $k$ (in our notation $k_{-}$ denotes the most
recent approximation to $k$).  For the linearization of the coupled
system \refeq{k:eps:coupled:dis}, we solve the problem
\beq
\tilde F_k + \tilde F_\epsilon = 0 \quad \forall \, (v,q)\in V_k \times V_\epsilon \ep
\label{k:eps:coupled:dis:lin}
\eeq

One often wants to linearize differently in segregated and coupled
formulations.  For example, in the coupled approach the $\epsilon$
term in \eqref{k:eq:LS:dis} may be rewritten as $\epsilon k/k$, with
the product $\epsilon k$ weighted according to
\beq
 (1-e_d)\epsilon_{-} k + e_d\epsilon k_{-},
      \quad e_d\in [0,1]\ep
\label{def:e_d}
\eeq
This yields a slightly different $\tilde F_k$ definition:
\begin{multline}
\tilde F_k
 \equiv
  - \int_{\Omega} \u\cdot\nabla k v_k \dif x
  - \int_{\Omega} \nu_{k-} \nabla k, \nabla v_k \dif x
  + \int_{\Omega} P_{k-} v_k \dif x
\\
  - \int_{\Omega} e_d \epsilon_{-} k/k_{-} + (1-e_d) \epsilon v_k \dif x
  -  \int_{\Omega} D_{-} v_k \dif x.
\label{k:eq:LS:dis:lin2}
\end{multline}
Our use of the underscore in variable names makes it particularly
easy to change linearizations.  If we want a term to be treated more
explicitly, or more implicitly, it is simply a matter of adding or
removing an underscore. For instance, {\fontsize{11pt}{11pt}\verb!f2*C_e2*e_*e/k_!},
corresponds to linearizing $f_2C_{e2}\epsilon^2/k$ as
$f_2C_{e2}\epsilon_{-}\epsilon/k_{-}$.  The whole term can be made
explicit and moved to the right-hand side of the linear system by
simply adding an underscore: {\fontsize{11pt}{11pt}\verb!f2*C_e2*e_*e_/k_!}.  On the contrary,
we could remove all the underscores to obtain a fully implicit term,
{\fontsize{11pt}{11pt}\verb!f2*C_e2*e*e/k!}.  This action would require that we use the
expression together with a full Newton method and a coupled formulation.

\paragraph{Newton methods}
A full Newton method for \refeq{k:eps:coupled:dis} involves a considerable
number of terms.  A modified Newton approach may be preferable, where
we linearize some terms as in the Picard strategy above and use a
Newton method to deal with the remaining nonlinear terms. For example,
previous iteration values can be used for $\nu_k$ while the $\nabla
k\cdot\nabla k$ factor in $D$ can be kept nonlinear.  A Newton method for
\refeq{k:eps:coupled:dis:lin} can also be formulated analogously to the
case where the $k$ and $\epsilon$ equations are solved in a segregated
manner.  In the implementation, we can specify the full nonlinear forms
$F_k$ and $F_\epsilon$, or we can do some manual Picard-type linearization
and then request automatic computation of the Jacobian.

It will turn out that different schemes can be tested easily using the
symbolic differentiation features of the form language UFL.  The Jacobian
for Newton methods will not need to be derived by hand, thereby avoiding a
process which is tedious and error-prone.  The details will be exemplified
in code extracts in the following section.

\section{Software design and implementation}
\label{sec:design}

Given a mathematical model, we propose to always distinguish between
code specific to a certain flow problem under investigation, code
responsible for solving the entire system of equations in a given
model and code responsible for solving each subsystem (some
PDEs, a single PDE or a term in a PDE) that makes up the
entire model  Here we refer to the first code
segment as a \emph{problem} class, the second as a \emph{solver} class
and the third as a \emph{scheme} class.  The problem code basically
defines the input to the solver and asks for a solution, while the
solver defines the complete PDE model in terms of a collection of
scheme objects and associated unknown functions.  The solver asks the
some of the scheme objects to set up and solve various parts of the
overall PDE model, while other scheme objects may compute quantities
derived from the primary unknowns, such as the strain rate tensor and
the turbulent viscosity. This design approach applies to Navier-Stokes
solvers, RANS models and in fact any model consisting of a system of
PDEs.

\subsection{Parameters}
\label{sec:parameters}

Flexible software frameworks for computational science normally involve a
large number of parameters that users can set. This is particularly true
for turbulent flows. Here we assume that each class, \emph{problem},
\emph{solver} or \emph{scheme}, creates its own {\fontsize{11pt}{11pt}\texttt{self.prm}} object
that is a dictionary of necessary parameters.  To look up a parameter,
say {\fontsize{11pt}{11pt}\texttt{order}}, one writes {\fontsize{11pt}{11pt}\texttt{self.prm['order']}}.  The parameter
dictionaries may also be nested and contain other dictionaries, where
appropriate. The user can operate the parameter pool directly, through
code, command-line options, a GUI or a web interface. Each class has
its own parameter dictionary that contains default values that may be
overloaded by the user.
\subsection{Navier-Stokes solvers}
\label{sec:NS:framework}
\subsubsection{Creating a solver}
\label{sec:NS:solver:usage}

In the context of laminar flow, {\fontsize{11pt}{11pt}\texttt{NSSolver}} and {\fontsize{11pt}{11pt}\texttt{NSProblem}} serve
as superclasses for the \emph{problem} and \emph{solver}, respectively.
Studying a specific flow case is a matter of creating a problem class,
say {\fontsize{11pt}{11pt}\texttt{MyProblem}}, by subclassing {\fontsize{11pt}{11pt}\texttt{NSProblem}} to inherit common
code and supplying at least four key methods: {\fontsize{11pt}{11pt}\texttt{mesh}} for returning
the (initial) finite element mesh to be used, {\fontsize{11pt}{11pt}\texttt{boundaries}} for
returning a list of different boundary types for the flow (wall,
inlet, outlet, periodic), {\fontsize{11pt}{11pt}\verb!body_force!} for specifying $\f$ and
{\fontsize{11pt}{11pt}\verb!initial_velocity_pressure!} for returning an initial guess of $\u$
and $p$ for the iterative solution approach. This guess can be a formula
({\fontsize{11pt}{11pt}\texttt{Expression}}), or perhaps computed elsewhere.  In addition, the
user must provide a {\fontsize{11pt}{11pt}\texttt{parameters}} dictionary with values for various
parameters in the simulation.  Important parameters are the polynomial
degree of the velocity and pressure fields, the solver type, the mesh
resolution, the Reynolds number and a scheme identifier.  The classes
{\fontsize{11pt}{11pt}\texttt{NSSolver}} and {\fontsize{11pt}{11pt}\texttt{NSProblem}} are located in Python modules with
the same name. The
default parameter dictionaries are defined in these modules.

A sample of the code needed to solve a flow problem with a coupled solver
may look as follows:
\begin{Verbatim}[fontsize=\fontsize{11pt}{11pt},tabsize=8,baselinestretch=1.0]
import cbc.rans.nsproblems as nsproblems
import cbc.rans.nssolvers as nssolvers

class MyProblem(nsproblems.NSProblem):
    ...

nsproblems.parameters.update(Nx=10, Ny=10, Re=100)
problem = MyProblem(nsproblems.parameters)
nssolvers.parameters = recursive_update(
          nssolvers.parameters, degree=dict(velocity=2,
          pressure=1), scheme_number=dict(velocity=1))
solver  = nssolvers.NSCoupled(problem, nssolvers.parameters)
solver.setup()
problem.solve(max_iter=20, max_err=1E-4)
plot(solver.u_); plot(solver.p_)
\end{Verbatim}
\noindent

\begin{figure}
  \center{\includegraphics[width=0.9\textwidth]{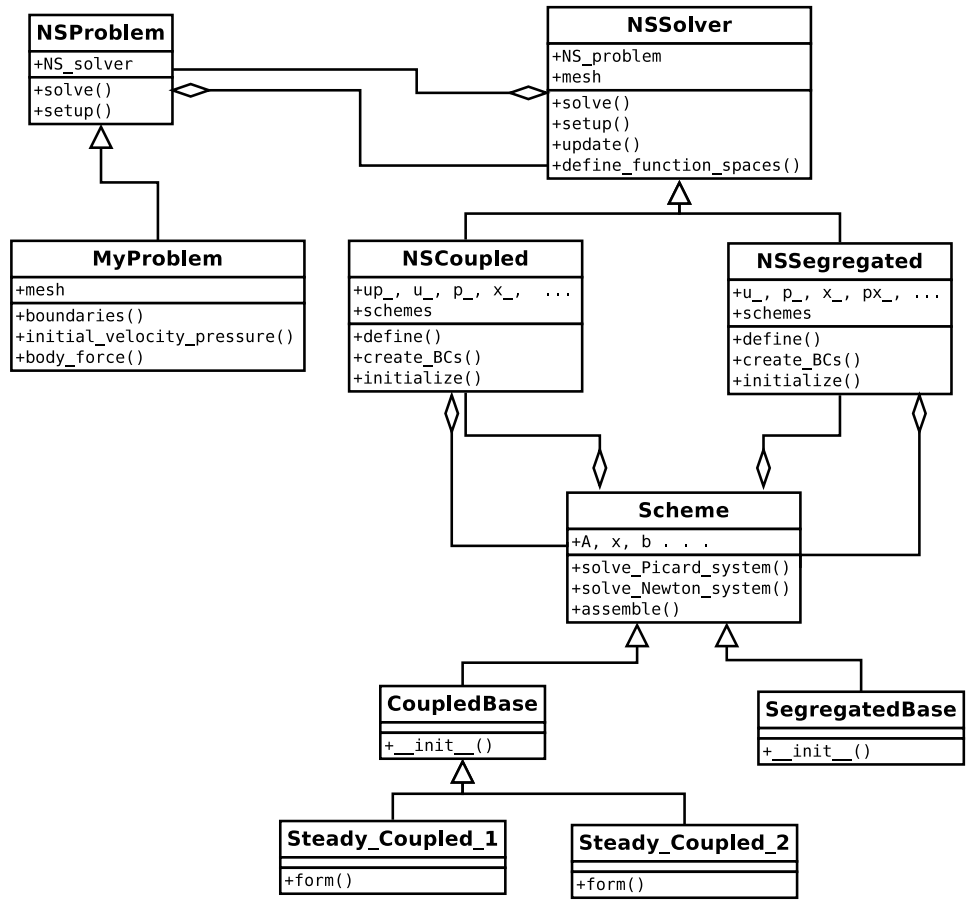}}
  \caption{UML sketch of some problem, solver and scheme classes (with
  some of their methods and attributes) for laminar flow modeled by the
  Navier-Stokes equations. The arrows with triangular heads represent
  classes derived from the classes that they point towards. The arrows
  with diamond heads indicate that the class pointed to is part of the
  class that is pointing. For example, an {\fontsize{11pt}{11pt}\texttt{NSSolver}} class contains
  a reference to an {\fontsize{11pt}{11pt}\texttt{NSProblem}} class and an {\fontsize{11pt}{11pt}\texttt{NSProblem}} class
  contains a reference to an {\fontsize{11pt}{11pt}\texttt{NSSolver}} class.}
\label{fig:NShier}
\end{figure}

Any computed quantity ($\u$, $p$, $\strainr$, $\int_\Omega\nabla\cdot\u\,
dx$, etc.)  is stored in the solver. The methods {\fontsize{11pt}{11pt}\texttt{setup}}
and {\fontsize{11pt}{11pt}\texttt{solve}} are general methods, normally inherited from the
superclass (note that the {\fontsize{11pt}{11pt}\texttt{setup}} method in an {\fontsize{11pt}{11pt}\texttt{NSProblem}}
class is called automatically by {\fontsize{11pt}{11pt}\texttt{setup}} in an {\fontsize{11pt}{11pt}\texttt{NSSolver}}
class).  The relationships between problem and solver classes are
outlined in the brief (and incomplete) Unified Modeling Language (UML)
diagram in Figure~\ref{fig:NShier}.  Note the introduction of a third
superclass {\fontsize{11pt}{11pt}\texttt{Scheme}}, designed to hold all information relevant to
the assembly and solve of one specific variational form. Subclasses in
the {\fontsize{11pt}{11pt}\texttt{Scheme}} hierarchy define one or more variational forms for
(parts of) PDE problems, assemble associated linear systems, and solve
these systems. Some forms arise in many PDE problems and collecting such
forms in a common library, together with assembly and solve functionality,
makes the forms reusable across many PDE solvers. This is the rationale
behind the {\fontsize{11pt}{11pt}\texttt{Scheme}} hierarchy.

A particular feature of classes in the {\fontsize{11pt}{11pt}\texttt{Scheme}} hierarchy is the
ease of avoiding assembly, and optimizing solve steps, if possible.
For example, if a particular form is constant in time, the {\fontsize{11pt}{11pt}\texttt{Scheme}}
subclass can easily assemble the associated matrix or vector only
once. If the form is also shared among PDE solvers, the various solvers
will automatically take advantage of only a single assembly operation.
Similarly, for direct solution methods for linear systems, the matrix
can be factored only once. Such optimization is of course dependent on
the discretization and linearization of the PDEs, which are details
that are defined by classes in the {\fontsize{11pt}{11pt}\texttt{Scheme}} hierarchy. Solvers can
then use {\fontsize{11pt}{11pt}\texttt{Scheme}} classes to compose the overall discretization and
solution strategy for a PDE or a system of PDEs.

Two solver classes are currently part of the {\fontsize{11pt}{11pt}\texttt{NSSolver}} hierarchy,
as depicted in Figure~\ref{fig:NShier}. {\fontsize{11pt}{11pt}\texttt{NSCoupled}} defines
function spaces and sets up common code (solution functions) for coupled
solvers, whereas {\fontsize{11pt}{11pt}\texttt{NSSegregated}} performs this task for solvers that
decouple the velocity from the pressure (e.g., fractional step methods).

The relationship between problem, solver, and scheme will be discussed
further in the remainder of this section.

\subsubsection{Solver classes}
\label{sec:NS:problem:solver}

Any problem class has a solver ({\fontsize{11pt}{11pt}\texttt{NSSolver}} subclass), and any solver
class has a reference back to the problem class.  It may be necessary
for several problem classes to share the same solver.  The key action
is calling {\fontsize{11pt}{11pt}\texttt{problem.solve}}, where the default implementation
in the superclass just calls the solver's {\fontsize{11pt}{11pt}\texttt{solve}} function. A
problem-specific version of {\fontsize{11pt}{11pt}\texttt{solve}} can alternatively be defined in
the user's problem class.

The {\fontsize{11pt}{11pt}\texttt{setup}} method in the class {\fontsize{11pt}{11pt}\texttt{NSSolver}} performs five important
initialization steps: extracting the mesh from the problem class, defining
function spaces, defining variational forms, initialization of velocity
and pressure functions and defining boundary conditions.  The definition
of function spaces and forms is done in methods that must normally be
overridden in subclasses, since these steps are usually tightly connected
to the numerical method used to solve the equations. Here is an example of
defining function spaces for $\u$, $p$, the compound function $(\u, p)$,
as well as a tensor function space for computing the strain rate tensor:
\begin{Verbatim}[fontsize=\fontsize{11pt}{11pt},tabsize=8,baselinestretch=1.0]
    def define_function_spaces(self):
        u_degree = self.prm['degree']['velocity'']
        p_degree = self.prm['degree']['pressure']

        self.V = VectorFunctionSpace(self.mesh, 'Lagrange', u_degree)
        self.Q = FunctionSpace(self.mesh, 'Lagrange', p_degree)
        self.VQ = self.V*self.Q  # mixed element

        # Symmetric tensor function space for strain rates Sij
        d = self.mesh.geometry().dim()  # space dim.
        symmetry = dict(((i,j), (j,i)) for i in range(d) \
                        for j in range(d) if i > j)
        self.S = TensorFunctionSpace(self.mesh, 'Lagrange', u_degree,
                                     symmetry=symmetry)
\end{Verbatim}
\noindent
We will sometimes show code like the above without further explanation.
The purpose is to outline possibilities and to provide a glimpse of the
size and nature of the code needed to realize certain functionality in
FEniCS and Python.

Consider the coupled numerical method from Section~\ref{NS:numerics} for
the Navier-Stokes equations, combined with Picard or Newton iteration
and under-relaxation.  We need finite element functions for
the most recently computed approximations $\u_{-}$ and $p_{-}$, named
{\fontsize{11pt}{11pt}\verb!self.u_!} and {\fontsize{11pt}{11pt}\verb!self.p_!} in the code.  Because we wish to solve a
nonlinear system for $(\u, p$), there is a need for the compound function
{\fontsize{11pt}{11pt}\verb!self.up_!} with a vector $x_{-}$ ({\fontsize{11pt}{11pt}\verb!self.x_!}) of degrees of
freedom. This vector should share storage with the vectors of $\u_{-}$ and
$p_{-}$. The update of {\fontsize{11pt}{11pt}\verb!self.x_!} is based on relaxing the solution
of the linear system with the old value of {\fontsize{11pt}{11pt}\verb!self.x_!}.  Taking into
account that $\u$ and $p$ are vector and scalar functions, respectively,
normally approximated by different types of finite elements, it is not
trivial to design a clean code (especially not in C and Fortran 77,
which are the dominant languages in the CFD).  There is, fortunately,
convenient support for working with functions and their vectors on
individual and mixed spaces in FEniCS. A typical initialization of data
structures is a shared effort between the superclass and the derived
solver class. The superclass is responsible for collecting all relevant
information from the problem, whereas the derived class initializes
solver specific functions and hooks up with appropriate schemes:
\begin{Verbatim}[fontsize=\fontsize{11pt}{11pt},tabsize=8,baselinestretch=1.0]
class NSSolver:
    ...
    def setup(self):
        self.NS_problem.setup(self)
        self.mesh = self.NS_problem.mesh
        self.define_function_spaces()
        self.u0_p0 = self.NS_problem.initial_velocity_pressure()
        self.boundaries = self.NS_problem.boundaries()
        self.f = self.NS_problem.body_force()
        self.nu = Constant(self.NS_problem.prm['viscosity'])

class NSCoupled(NSSolver)
    ...
    def setup(self):
        NSSolver.setup(self)
        VQ = self.VQ
        (self.v, self.q) = TestFunctions(VQ)
        self.up = TrialFunction(VQ)
        (self.u, self.p) = ufl.split(self.up)
        self.bcs = self.create_BCs(self.boundaries)
        self.up_ = Function(VQ)
        self.u_, self.p_ = self.up_.split()
        self.x_ = self.up_.vector()
        self.initialize(self.u0_p0.vector())
        self.schemes = {'NS': None, 'parameters': []}
        self.define()
\end{Verbatim}
\noindent
Creating {\fontsize{11pt}{11pt}\verb!self.up_!} as a $(\u, p)$ function on $V\times Q$ and then
splitting this compound function into parts on $V$ and $Q$, gives two
\emph{references} (``pointers'' in C-style terminology): {\fontsize{11pt}{11pt}\verb!self.u_!} to
$\u_{-}$ and {\fontsize{11pt}{11pt}\verb!self.p_!} to $p_{-}$.  Whenever we update {\fontsize{11pt}{11pt}\verb!self.u_!}
or {\fontsize{11pt}{11pt}\verb!self.p_!}, {\fontsize{11pt}{11pt}\verb!self.up_!} is also updated, and vice versa.
Similarly, updating {\fontsize{11pt}{11pt}\verb!self.x_!} \emph{in-place} updates the values of
the compound function {\fontsize{11pt}{11pt}\verb!self.up_!} and its parts {\fontsize{11pt}{11pt}\verb!self.u_!} and
{\fontsize{11pt}{11pt}\verb!self.p_!}, since memory is shared.  That is, we can work with $\u$
or $p$ or $(\u, p)$, or their corresponding degrees of freedom vectors
interchangeably, according to what is the most appropriate abstraction
for a given operation.  The generalization to a more complicated system
of vector and scalar PDEs is straightforward.

The {\fontsize{11pt}{11pt}\texttt{self.nu}} variable deserves a comment. For
laminar flow, {\fontsize{11pt}{11pt}\texttt{self.nu}} will typically be a
{\fontsize{11pt}{11pt}\verb!Constant(self.NS_problem.prm['viscosity'])!}, but in turbulence
computations {\fontsize{11pt}{11pt}\texttt{self.nu}} must refer to this constant plus a finite
element representation of $\nu_T$ (see equation~\eqref{eq:RANS:m}).  This
is accomplished by a simple (re)assignment in the turbulent case. Computer
languages with static typing would here need some parameterization
of the type, when it changes from {\fontsize{11pt}{11pt}\texttt{Constant}} to {\fontsize{11pt}{11pt}\texttt{Constant}}
+ {\fontsize{11pt}{11pt}\texttt{Function}}.  Normally, this requires nontrivial object-oriented or
generative programming in C++, but dynamic typing in Python makes
an otherwise complicated technical problem trivial. Especially in
PDE solver frameworks, new logical combinations are needed, as is the
ability to let variables point to new objects since this leads to simple
and compact code.  The corresponding code in C++, Java, or C\# would
usually introduce extra classes to help ``simulate'' flexible references,
resulting in frameworks with potentially a large number of classes.

\subsubsection{Iteration schemes}
\label{sec:NS:iterschemes}

Subclasses of the {\fontsize{11pt}{11pt}\texttt{Scheme}} class hierarchy implement specific
linearized variational forms that can be combined in solver classes
to implement various discretizations of the governing system of PDEs.
As mentioned, reuse of common variational forms, their matrices and
preconditioners, as well as encapsulation of optimization tricks are the
primary reasons for introducing the {\fontsize{11pt}{11pt}\texttt{Scheme}} hierarchy.  Here is one
class for the variational forms associated with a fully coupled NS solver:
\begin{Verbatim}[fontsize=\fontsize{11pt}{11pt},tabsize=8,baselinestretch=1.0]
class CoupledBase(Scheme):
    def __init__(self, solver, unknown):
        Scheme.__init__(self, solver, unknown, ...)
        form_args = vars(solver).copy()
        if self.prm['iteration_type'] == 'Picard':
            F = self.form(**form_args)
            self.a, self.L = lhs(F), rhs(F)
        elif self.prm['iteration_type'] == 'Newton':
            form_args['u_'], form_args['p_'] = solver.u, solver.p
            up_, up = unknown, solver.up
            F = self.form(**form_args)
            F_ = action(F, function=up_)
            J_ = derivative(F_, up_, up)
            self.a, self.L = J_, -F_
\end{Verbatim}
\noindent
Subclasses of {\fontsize{11pt}{11pt}\texttt{Scheme}} hold the forms {\fontsize{11pt}{11pt}\texttt{a}} and {\fontsize{11pt}{11pt}\texttt{L}} that are
needed for forming the linear system associated with the variational
form represented by the class. Typically, a method {\fontsize{11pt}{11pt}\texttt{form}} (in a
subclass of {\fontsize{11pt}{11pt}\texttt{CoupledBase}}) defines this variational form, here
stored in the {\fontsize{11pt}{11pt}\texttt{F}} variable, and then the {\fontsize{11pt}{11pt}\texttt{a}} and {\fontsize{11pt}{11pt}\texttt{L}} parts
are extracted. Note that a full Newton method is easily formulated,
thanks to UFL's support for automatic differentiation. First, we define
the nonlinear variational form {\fontsize{11pt}{11pt}\texttt{F}} by substituting the variable
{\fontsize{11pt}{11pt}\verb!u_!} in the scheme by the trial function {\fontsize{11pt}{11pt}\texttt{solver.u}} (similar
for the pressure). Second, the right-hand side is generated by applying
the nonlinear form {\fontsize{11pt}{11pt}\texttt{F}} as an action on the most recently computed
unknown function (i.e., the trial function is replaced by {\fontsize{11pt}{11pt}\verb!up_!},
which is {\fontsize{11pt}{11pt}\verb!solver.up_!}). Then we can compute the Jacobian of the
nonlinear form in one line.

Besides defining and storing the forms {\fontsize{11pt}{11pt}\texttt{self.a}} and {\fontsize{11pt}{11pt}\texttt{self.L}}, a
scheme class also assembles the associated matrix {\fontsize{11pt}{11pt}\texttt{self.A}} and vector
{\fontsize{11pt}{11pt}\texttt{self.b}}, and solves the system for the solution {\fontsize{11pt}{11pt}\texttt{self.x}}.
The latter variable simply refers to the vector storage of the
{\fontsize{11pt}{11pt}\verb!solver.up_!} {\fontsize{11pt}{11pt}\texttt{Function}}. That is, the solver is responsible
for creating storage for the primary unknowns and derived quantities,
while scheme classes create storage for the matrix and right-hand side
associated with the solution of the equations implied by the variational forms.

Subclasses of {\fontsize{11pt}{11pt}\texttt{CoupledBase}} provide the exact formula for the
variational form through the {\fontsize{11pt}{11pt}\texttt{form}} method. Here is an example of
a fully implicit scheme:
\begin{Verbatim}[fontsize=\fontsize{11pt}{11pt},tabsize=8,baselinestretch=1.0]
class Steady_Coupled_1(CoupledBase):
    def form(self, u, v, p, q, u_, nu, f, **kwargs):
        return inner(v, dot(grad(u), u_))*dx \
            + nu*inner(grad(v), grad(u)+grad(u).T)*dx - inner(v, f)*dx \
            - inner(div(v), p)*dx - inner(q, div(u))*dx
\end{Verbatim}
\noindent
The required arguments are passed to {\fontsize{11pt}{11pt}\texttt{form}} as a namespace dictionary
containing all variables in the
solver (see the constructor of {\fontsize{11pt}{11pt}\texttt{CoupledBase}} where {\fontsize{11pt}{11pt}\texttt{form}} is
called).  Alternatively, we may list only those variables that are needed as
arguments to the {\fontsize{11pt}{11pt}\texttt{form}} method, at the cost of extensive writing
if numerous parameters are needed in the form (as in RANS models).
Note that the {\fontsize{11pt}{11pt}\texttt{**kwargs}} argument absorbs
all the extra uninteresting variables in the call that do not match the
names of the positional arguments.  Yet, there is no additional overhead
involved, because the {\fontsize{11pt}{11pt}\texttt{**kwargs}} dictionary is simply a pointer to
the solver's namespace.

Picard and Newton variants can both employ the form shown above -- the
difference is simply the {\fontsize{11pt}{11pt}\verb!u_!} argument ($\u_{-}\cdot\nabla\u$
versus $\u\cdot\nabla\u$). Setting the {\fontsize{11pt}{11pt}\verb!u_!} variable in the
namespace dictionary {\fontsize{11pt}{11pt}\verb!form_args!} to {\fontsize{11pt}{11pt}\verb!solver.u!} instead of
{\fontsize{11pt}{11pt}\verb!solver.u_!}, makes the first term evaluate to the nonlinear form
{\fontsize{11pt}{11pt}\texttt{inner(v, dot(grad(u), u))*dx}}.

An explicit scheme, utilizing only old velocities in the convection term,
is implemented similarly:
\begin{Verbatim}[fontsize=\fontsize{11pt}{11pt},tabsize=8,baselinestretch=1.0]
class Steady_Coupled_2(CoupledBase):
    def form(self, u, v, p, q, u_, nu, f, **kwargs):
       if type(solver.nu) is Constant and \
           self.prm['iteration_type'] == 'Picard':
           self.prm['reassemble_lhs'] = False

       return inner(v, dot(grad(u_), u_))*dx \
           + nu*inner(grad(v), grad(u)+grad(u).T)*dx - inner(v, f)*dx \
           - inner(div(v), p)*dx - inner(q, div(u))*dx
\end{Verbatim}
\noindent
The convective term for the explicit scheme is different from that for
the implicit scheme, but we also flag that in a Picard iteration, for
constant viscosity,
the coefficient matrix does not change since the convective term only
contributes to the right-hand side, implying that reassembly can
be avoided. Such optimizations are key features of classes in the
{\fontsize{11pt}{11pt}\texttt{Schemes}} hierarchy.

In the real implementation
of our framework, the convective term is evaluated by a
separate method where one can choose between several alternative
formulations of this term. Also, stabilization terms, like shown in
\eqref{eq:NS:varform:coupled_stabilized}, can be added in the
{\fontsize{11pt}{11pt}\texttt{form}} method.

The solver class, which one normally would assign the task of defining
variational forms, now refers to subclass(es) of {\fontsize{11pt}{11pt}\texttt{Scheme}} for
defining appropriate forms and also for assembling matrices and solving
linear systems. The solver class holds the system of equations, and each
individual equation is represented as a scheme class.  A coupled solver
adds the necessary schemes to a {\fontsize{11pt}{11pt}\texttt{schemes}} dictionary as part of the
setup procedures:
\begin{Verbatim}[fontsize=\fontsize{11pt}{11pt},tabsize=8,baselinestretch=1.0]
class NSCoupled(NSSolver):
    ...
    def define(self):
        # Define a Navier-Stokes scheme
        classname = self.prm['time_integration'] + '_Coupled_' + \
                    str(self.prm['scheme_number']['velocity'])
        self.schemes['NS'] = eval(classname)(self, self.up_)
\end{Verbatim}
\noindent
User-given parameters are used to construct the appropriate name of the
subclass of {\fontsize{11pt}{11pt}\texttt{Scheme}} that defines the relevant form.  With {\fontsize{11pt}{11pt}\texttt{eval}}
we can turn this name into a living object, without the usual {\fontsize{11pt}{11pt}\texttt{if}}
or {\fontsize{11pt}{11pt}\texttt{case}} statements in factory functions
that would be necessary in C, C++, Fortran, and Java.
\subsubsection{Derived quantities}
\label{sec:NS:dq}

Derived quantities, such as the strain rate and stress tensors, can be
computed once $\u$ and $p$ are available.  For a low-Reynolds turbulence
model, Table~\ref{tab:lowre} lists numerous quantities that must be
derived from the primary unknowns in the system of PDEs. Some of the
derived quantities can be computed from the primary unknowns without
any derivatives, e.g., $\nu_T=C_ \mu f_ \mu k^2/\epsilon$ with $f_ \mu$
being an exponential function of $Re_T\sim k^2/\epsilon$.  One can either
project the expression of $\nu_T$ onto a finite element space or one can
compute the degrees of freedom of $\nu_T$ directly from the degrees of
freedom of $k$ and $\epsilon$.  Other derived quantities, such as $D$ in
Table~\ref{tab:lowre}, involve derivatives of the primary unknowns. These
derivatives are discontinuous across cell facets, and when
needed in some variational form, we can either use the quantity's form
as it is, or we may choose to first project the quantity onto a finite
element space of continuous functions and then use it in other contexts.

To effectively define and work with the large number of derived quantities
in RANS models, we need a flexible code construction where we essentially
write the formula defining a derived quantity $Q$ and then choose
between three ways of utilizing the formula: we may (i) \emph{project}
$Q$ onto a space $V$, (ii) \emph{use the formula} for $Q$ directly in
some variational form, or (iii) \emph{compute the degrees of freedom}
of $Q$, by applying the formula to each individual degree of freedom,
for efficiently creating a finite element function of $Q$.  A class
{\fontsize{11pt}{11pt}\texttt{DerivedQuantity}} is designed to hold the definition of a derived
quantity and to apply it in one of the three aforementioned ways. In
some solver class (like {\fontsize{11pt}{11pt}\texttt{NSCoupled}}) we can define the computation
of a derived quantity, say the strain rate tensor $\strainr$, by
\begin{Verbatim}[fontsize=\fontsize{11pt}{11pt},tabsize=8,baselinestretch=1.0]
Sij = DerivedQuantity(solver=self, name='Sij', space=self.S,
      formula='strain_rate(u_)', namespace=ns, apply='project')
\end{Verbatim}
\noindent
The formula for $\strainr$ makes use of a Python function
\begin{Verbatim}[fontsize=\fontsize{11pt}{11pt},tabsize=8,baselinestretch=1.0]
def strain_rate(u):
    return 0.5*(grad(u) + grad(u).T)
\end{Verbatim}
\noindent
Alternatively, the {\fontsize{11pt}{11pt}\texttt{formula}} argument could be the expression
inside the {\fontsize{11pt}{11pt}\verb!strain_rate!} function (with {\fontsize{11pt}{11pt}\texttt{u}} replaced
by {\fontsize{11pt}{11pt}\verb!u_!}). We may nest functions for
defining derived quantities, e.g., the stress tensor could be defined
as {\fontsize{11pt}{11pt}\verb!formula='stress(u_, p_, nu)'!} where
\begin{Verbatim}[fontsize=\fontsize{11pt}{11pt},tabsize=8,baselinestretch=1.0]
def stress(u, p, nu):
    d = u.cell().d  # no of space dimensions
    return 2*nu*strain_rate(u) - p*Identity(d)
\end{Verbatim}
\noindent
The {\fontsize{11pt}{11pt}\texttt{namespace}} argument must hold a namespace dictionary in
which the string {\fontsize{11pt}{11pt}\texttt{formula}} is going to be evaluated by {\fontsize{11pt}{11pt}\texttt{eval}}.
It means, in the present example, that {\fontsize{11pt}{11pt}\texttt{ns}} must be a dictionary
defining {\fontsize{11pt}{11pt}\verb!u_!}, {\fontsize{11pt}{11pt}\verb!strain_rate!}, and other objects that are
needed in the formula for the derived quantity. A quick construction
of a common namespace for most purposes is to let {\fontsize{11pt}{11pt}\texttt{ns}} be the merge
of {\fontsize{11pt}{11pt}\texttt{vars(self)}} (all attributes in the solver) and {\fontsize{11pt}{11pt}\texttt{globals()}}
(all the global functions and variables in the solver module).

The {\fontsize{11pt}{11pt}\texttt{apply}} argument specifies how the formula is applied: for
projection ({\fontsize{11pt}{11pt}\texttt{'project'}}), direct computations of degrees of
freedom ({\fontsize{11pt}{11pt}\verb!'compute_dofs'!}), or plain use of the formula
({\fontsize{11pt}{11pt}\verb!'use_formula'!}). Other arguments are optional and
may specify how to solve the linear system arising in projection, how
to under-relax the projected quantity, etc. When this information is
lacking, the {\fontsize{11pt}{11pt}\texttt{DerivedQuantity}} class looks up missing information
in the parameters ({\fontsize{11pt}{11pt}\texttt{prm}}) dictionary in the solver class.

A formula for a derived quantity may involve previously defined
quantities. Therefore, since ordering is key, a solver will typically
collect its definitions of derived quantities in an ordered list.

A {\fontsize{11pt}{11pt}\texttt{DerivedQuantity}} object is a special kind of a {\fontsize{11pt}{11pt}\texttt{Scheme}} object,
and therefore naturally derives from {\fontsize{11pt}{11pt}\texttt{Scheme}}. The inner workings
depend on quite advanced Python coding, but yield great flexibility.
The fundamental idea is to specify the formula as a \emph{string}, and not
a UFL expression, because such a string can be evaluated by {\fontsize{11pt}{11pt}\texttt{eval}}
in different namespaces, yielding different results. Say we have a
{\fontsize{11pt}{11pt}\texttt{DerivedQuantity}} object with some formula {\fontsize{11pt}{11pt}\texttt{'k**2'}}. With a
namespace {\fontsize{11pt}{11pt}\texttt{ns}} where {\fontsize{11pt}{11pt}\texttt{k}} is tied to an object {\fontsize{11pt}{11pt}\texttt{k}} of type
{\fontsize{11pt}{11pt}\texttt{TrialFunction}}, {\fontsize{11pt}{11pt}\texttt{ns['k'] = k}}, the call {\fontsize{11pt}{11pt}\texttt{eval(formula, ns)}} will
turn the string into a UFL expression where {\fontsize{11pt}{11pt}\texttt{k}} is an unknown finite
element function.  On the other hand, with {\fontsize{11pt}{11pt}\verb!ns['k'] = k_!}, {\fontsize{11pt}{11pt}\verb!k_!}
being an already computed finite element function, the {\fontsize{11pt}{11pt}\texttt{eval}} call
turns {\fontsize{11pt}{11pt}\texttt{'k**2'}} into {\fontsize{11pt}{11pt}\verb!'k_**2!}, which yields a known right-hand
side in a projection or a known source term in a variational form.
Moreover, {\fontsize{11pt}{11pt}\texttt{ns['k'] = k.vector().array()}} associates the variable
{\fontsize{11pt}{11pt}\texttt{k}} in the formula with its array of the degrees of freedom, and
the {\fontsize{11pt}{11pt}\texttt{eval}} call will then lead to squaring this array. The result can be
inserted into the vector of degrees of freedom of a finite element field
to yield a more efficient computation of the field than the
projection approach.

Derived quantities that are projected may need to overload the default
boundary conditions through the {\fontsize{11pt}{11pt}\verb!create_BCs!} method.
The {\fontsize{11pt}{11pt}\texttt{DerivedQuantity}} class is by default set to
enforce assigned boundary conditions on walls, whereas a subclass
{\fontsize{11pt}{11pt}\verb!DerivedQuantity_NoBC!} does not. The latter is in fact used by the
implemented shear stress {\fontsize{11pt}{11pt}\verb!Sij!}, since the velocity gradient on a
wall in general will be unknown.

Especially in complex mathematical models with a range of quantities
that are defined as formulae involving the primary unknowns,
the {\fontsize{11pt}{11pt}\texttt{DerivedQuantity}} class helps to shorten application code
considerably and at the same time offer flexibility with respect to
explicit versus implicit treatment of formulae, projection of quantities
for visualization, etc.

\subsubsection{Solution of linear systems}
\label{sec:NS:linsys}

The {\fontsize{11pt}{11pt}\texttt{Scheme}} classes are responsible for solving the linear
system associated with a form. Since the Picard and Newton methods have
different unknowns in the linear system ($\u$ and $p$ versus corrections
of $\u$ and $p$), a general solve method is provided for each of them. The
Picard version with under-relaxation reads
\begin{Verbatim}[fontsize=\fontsize{11pt}{11pt},tabsize=8,baselinestretch=1.0]
def solve_Picard_system(self, assemble_A, assemble_b):
    for name in ('A', 'x', 'b', 'bcs'):
        exec str(name + ' = self.' + name)  # strip off self.
    if assemble_A: self.assemble(A)
    if assemble_b: self.assemble(b)
    [bc.apply(A, b) for bc in bcs]  # boundary conditions modify A, b
    self.setup_solver(assemble_A, assemble_b)
    x_star = self.work
    x_star[:] = x[:]  # start vector for iterative solvers
    self.linear_solver.solve(A, x_star, b)
    # relax: x = (1-omega)*x + omega*x_star = x + omega*(x_star-x)
    omega = self.prm['omega']
    x_star.axpy(-1., x); x.axpy(omega, x_star)
    self.update()
    return residual(A, x, b), x_star
\end{Verbatim}
\noindent
Note how we first strip off the {\fontsize{11pt}{11pt}\texttt{self}} prefix (by loading
attributes into local variables) to make the code easier to read and
closer to the mathematical description. This trick is frequently used throughout
our software to shorten the distance between code and mathematical expressiveness.
The linear system is assembled only if the previously computed {\fontsize{11pt}{11pt}\texttt{A}}
or {\fontsize{11pt}{11pt}\texttt{b}} cannot be reused. Similarly, if {\fontsize{11pt}{11pt}\texttt{A}} can be reused, the
factorization or preconditioner in a linear solver can also be reused (the
{\fontsize{11pt}{11pt}\verb!setup_solver!} method will pass on such information to the linear
solver). After the linear solver has computed the solution {\fontsize{11pt}{11pt}\verb!x_star!},
the new vector of velocities and pressures, {\fontsize{11pt}{11pt}\texttt{x}}, is computed by
relaxation. For this purpose we use the classical ``axpy'' operation:
$y\leftarrow ax+y$ ($a$ is scalar, $x$ and $y$ are vectors). Since
``axpy'' is an efficient operation (carried out in, e.g., PETSc if that
is the chosen linear algebra backend for FEniCS), we rewrite the usual
relaxation update formula to fit with this operation. The in-place update
of {\fontsize{11pt}{11pt}\verb!x!} through the {\fontsize{11pt}{11pt}\texttt{axpy}} method is essential when {\fontsize{11pt}{11pt}\verb!x!}
has memory shared with several finite element functions, as explained
earlier. The returned values are the solution of one iteration, the
corresponding residual and the difference between the previous and the
new solution (reflected by {\fontsize{11pt}{11pt}\verb!x_star!} after its {\fontsize{11pt}{11pt}\texttt{axpy}} update).

The solution of a linear system arising in Newton methods requires a
slightly different function, because we
solve for a correction vector, and the residual is the right-hand side
of the system.
\begin{Verbatim}[fontsize=\fontsize{11pt}{11pt},tabsize=8,baselinestretch=1.0]
def solve_Newton_system(self, *args):
    for name in ('A', 'x', 'b', 'bcs'):
        exec str(name + ' = self.' + name)
    self.assemble(A)
    self.assemble(b)
    [bc.apply(A, b, x) for bc in bcs]
    dx =self. work  # more informative name
    dx.zero()
    self.linear_solver.solve(A, dx, b)
    x.axpy(self.prm['omega'], dx)
    self.update()
    return norm(b), dx
\end{Verbatim}
\noindent
The dummy arguments {\fontsize{11pt}{11pt}\texttt{*args}} are included in the call (but never
used) so that {\fontsize{11pt}{11pt}\verb!solve_Picard_system!} and {\fontsize{11pt}{11pt}\verb!solve_Newton_system!}
can be called with the same set of arguments. A simple wrapper function
{\fontsize{11pt}{11pt}\texttt{solve}} will then provide a uniform interface to either the Picard or
Newton version for creating and solving a linear system:
\begin{Verbatim}[fontsize=\fontsize{11pt}{11pt},tabsize=8,baselinestretch=1.0]
def solve(self, assemble_A=None, assemble_b=None):
    return eval('self.solve_%s_system' % self.prm['iteration_type'])\
               (assemble_A, assemble_b)
\end{Verbatim}
\noindent
With this {\fontsize{11pt}{11pt}\texttt{solve}} method, it is easy to write a general iteration
loop to reach a steady state solution. This loop is independent of
whether we use the Newton or Picard method, or how we avoid assembly
and reuse matrices and vectors:
\begin{Verbatim}[fontsize=\fontsize{11pt}{11pt},tabsize=8,baselinestretch=1.0]
def solve_nonlinear(scheme, max_iter=1, max_err=1e-8, update=None):
    j = 0; err = 1E+10
    scheme.info = {'error': (0,0), 'iter': 0}
    while err > max_err and j < max_iter:
        res, dx = scheme.solve(scheme.prm['reassemble_lhs'],
                               scheme.prm['reassemble_rhs'])
        j += 1
        scheme.info = {'error': (res, norm(dx)), 'iter': j}
        if scheme.prm['echo']: print scheme.info
        err = max(scheme.info['error'])
        if update: update()
    return scheme.info
\end{Verbatim}
\noindent
The {\fontsize{11pt}{11pt}\texttt{scheme}} object is a subclass of {\fontsize{11pt}{11pt}\texttt{Scheme}} that has the
{\fontsize{11pt}{11pt}\texttt{solve}} method listed above.  The {\fontsize{11pt}{11pt}\texttt{update}} argument is usually
some method in the {\fontsize{11pt}{11pt}\texttt{solver}} object that updates data structures
of interest, which could be some derived quantity (e.g., $\strainr$ and
$\nabla\cdot\u$). It can also be used to plot or save intermediate results
between iterations. Note that {\fontsize{11pt}{11pt}\texttt{scheme}} also has an {\fontsize{11pt}{11pt}\texttt{update}}
method that is called at the end of {\fontsize{11pt}{11pt}\verb!solve_Picard/Newton_system!}.
The {\fontsize{11pt}{11pt}\texttt{scheme.update}} method is often used to enforce additional control
over {\fontsize{11pt}{11pt}\verb!x!}, e.g., for {\fontsize{11pt}{11pt}\verb!k_!} by ensuring that it is always larger
than zero.

A useful Python feature is the ability to define new class attributes
whenever appropriate, and is used in the preceding snippet for storing
information about the iteration in {\fontsize{11pt}{11pt}\texttt{scheme.info}}.  This presents
the possibility of adding new components to a framework to dynamically
increase functionality. Simple code may remain simple, even when
extensions are required for more complex cases, since extensions can be
added when needed at run-time by other pieces of the software.

In a classical object-oriented C++ design, the stand-alone
{\fontsize{11pt}{11pt}\verb!solve_nonlinear!} function would naturally be a method in an NS
solver superclass. However, reuse of this generic iteration function
to solve other equations then forces those equations to have their
solvers as subclasses in the NS hierarchy. Also, the shown version of
{\fontsize{11pt}{11pt}\verb!solve_nonlinear!} is very simple, checking only the size of the norms
of the residual for convergence.  More sophisticated stopping criteria
can be implemented and added trivially.  Alternatively, a user may want a
tailored {\fontsize{11pt}{11pt}\verb!solve_nonlinear!} function. This is trivially accomplished,
whereas if the function were placed inside a class in a class hierarchy,
the user would need to subclass that class and override the function.
This approach connects the new function to a particular solver class,
while a stand-alone function can be combined with any solver class from
any solver hierarchy as long as the solver provides certain attributes and
methods.  The same flexibility can be achieved by generative programming
in C++ via templates.

The {\fontsize{11pt}{11pt}\texttt{solve}} methods in solver classes will typically make use
of {\fontsize{11pt}{11pt}\verb!solve_nonlinear!} or variants of it for performing the solve
operation.

\subsection{Reynolds-averaged Navier-Stokes models}
\label{sec:RANS:framework}

The class {\fontsize{11pt}{11pt}\texttt{NSSolver}} and its subclasses are designed to be used
without any turbulence model, but with the possibility of having a
variable viscosity.  Since RANS models are implemented separately in
their own classes, we need to decide on the relation between NS solver
classes and RANS solver classes.  There are three obvious approaches:
(1) let a RANS model be a subclass of an NS solver class; (2) let a RANS
model have a reference to an instance of an NS solver; or (3) let a RANS
solver only define and solve RANS equations, and then use a third class
to couple NS and RANS classes.  We want maximum flexibility in the sense
that any solution method for the NS equations can in principle be used
with any turbulence model. Approach~(1), with subclassing RANS models,
ties a RANS model to a particular NS class and thus limits flexibility.
With approaches (2) and (3), the user selects any RANS model and any NS
solution method. Since a RANS model is incomplete without an NS solver,
we prefer approach~(2).

\begin{figure}
  \center{\includegraphics[width=0.8\textwidth]{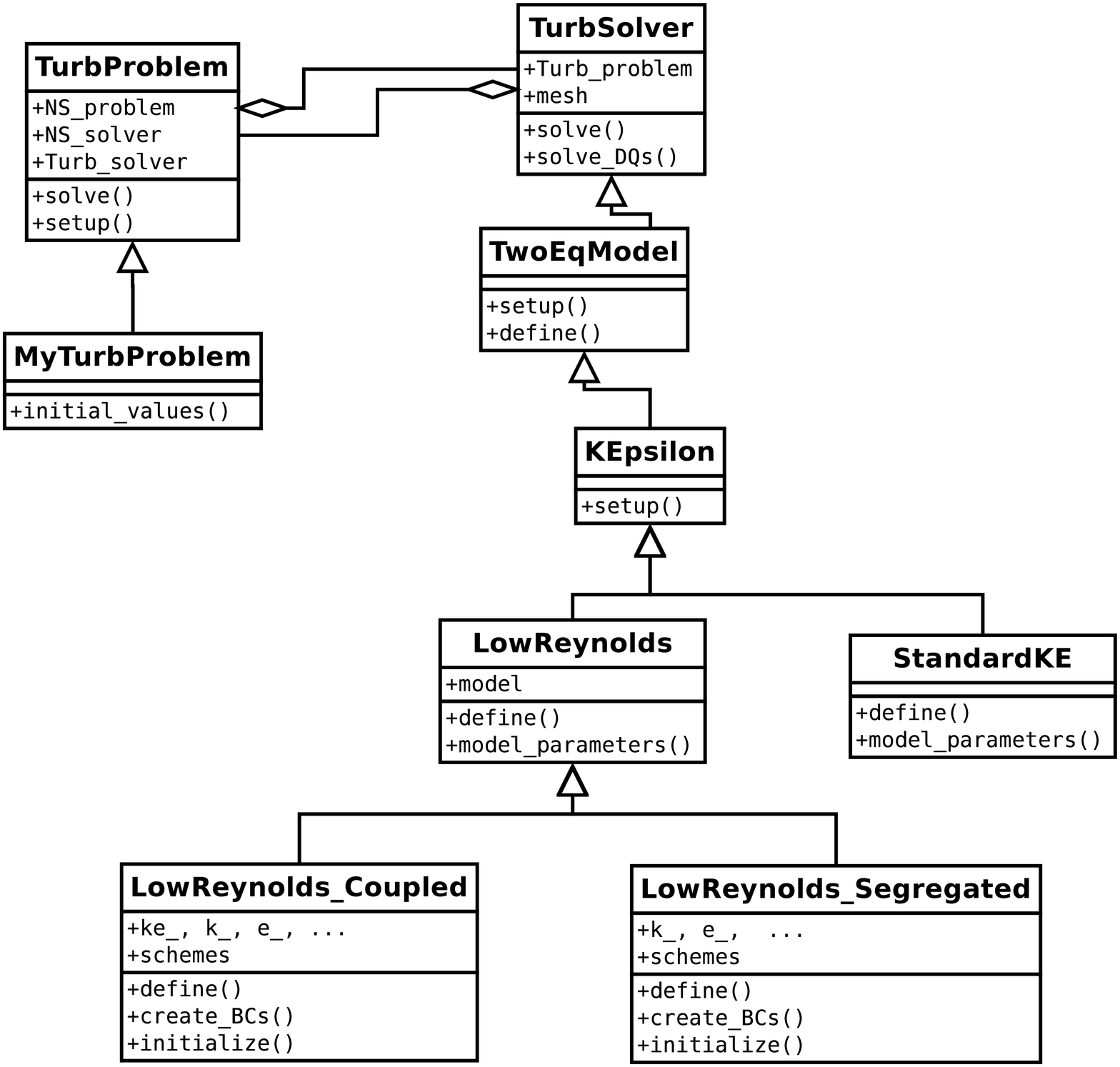}}
  \caption{Sketch of a few problem and solver classes (with some of
  their methods and attributes) for turbulent flow modeled by RANS.}
\label{fig:Turbhier}
\end{figure}

Mirroring the structure of the NS solver, RANS solvers have a superclass
{\fontsize{11pt}{11pt}\texttt{TurbSolver}}, while the {\fontsize{11pt}{11pt}\texttt{TurbProblem}} acts as superclass for the
turbulence problems.  Any turbulence problem contains an {\fontsize{11pt}{11pt}\texttt{NSProblem}}
class for defining the basic flow problem, plus parameters related to
turbulence PDEs and their solution methods. Figure~\ref{fig:Turbhier}
sketches the relationships between some of the classes to be discussed
in the text.

\subsubsection{Required functionality}
\label{sec:RANS:requirements}

RANS modeling poses certain numerical challenges that a software system
must be able to deal with in a flexible way. It must be easy to add new
PDEs and combine PDEs with various constitutive relations to form new
models or variations on classical ones.  Due to the nonlinearties in
turbulence PDEs, the degree of implicitness when designing an effective
and robust iteration method is critical. We wish to make switching between
implicit and explicit treatments of terms in an equation straightforward,
thereby offering complete control over the linearization procedure.
Key is the flexibility to construct schemes.  This is made possible
in part by automatic symbolic differentiation, which can be applied
selectively to different terms.
\subsubsection{Creating a turbulent flow problem}
\label{sec:RANS:problem}

Solving a turbulent flow problem is a matter of extending the code
example from Section~\ref{sec:NS:solver:usage}. We make use of the same
{\fontsize{11pt}{11pt}\texttt{MyProblem}} class for defining a mesh, etc., but for a turbulent flow
we might want to initialize the velocity and pressure differently. For
this purpose we overload the {\fontsize{11pt}{11pt}\verb!initial_velocity_pressure!} method
in {\fontsize{11pt}{11pt}\texttt{MyProblem}}.  A subclass of {\fontsize{11pt}{11pt}\texttt{TurbProblem}} must be defined to
set initial conditions for the turbulence equations and supply the solver
with the correct boundary values (the boundaries are already supplied by
{\fontsize{11pt}{11pt}\texttt{MyProblem}}).  The creation of problem and solver classes, and setting
of parameters through predefined dictionaries in the {\fontsize{11pt}{11pt}\texttt{cbc.rans}}
modules, may look as follows for a specific flow case:
\begin{Verbatim}[fontsize=\fontsize{11pt}{11pt},tabsize=8,baselinestretch=1.0]
import cbc.rans.nsproblems as nsproblems
import cbc.rans.nssolvers as nssolvers
import cbc.rans.turbproblems as turbproblems
import cbc.rans.turbsolvers as turbsolvers

class MyProblem(nsproblems.MyProblem):
    """Overload initialization of velocity/pressure."""
    def initial_velocity_pressure(self):
        ...

class MyProblemTurb(TurbProblem):
    ...

nsproblems.parameters.update(Nx=10, Ny=10)
turbproblems.parameters.update(Model='Chien', Re_tau=395.)
nsproblems.parameters.update(turbproblems.parameters)
NS_problem = MyProblem(nsproblems.parameters)
nssolvers.parameters = recursive_update(nssolvers.parameters,
          dict(degree=dict(velocity=2, pressure=1)))
NS_solver  = nssolvers.NSCoupled(NS_problem, nssolvers.parameters)
NS_solver.setup()

problem = MyTurbProblem(NS_problem, turbproblems.parameters)
turbsolvers.parameters.update(iteration_type='Picard', omega=0.6)
solver = turbsolvers.LowReynolds_Coupled(problem,
                                         turbsolvers.parameters)
solver.setup()
problem.solve(max_iter=10)
\end{Verbatim}
\noindent
\subsubsection{Turbulence model solver classes}
\label{sec:RANS:solver}

The {\fontsize{11pt}{11pt}\texttt{TurbSolver}} class has a relation to the {\fontsize{11pt}{11pt}\texttt{TurbProblem}} class
that mimics the relation between {\fontsize{11pt}{11pt}\texttt{NSSolver}} and {\fontsize{11pt}{11pt}\texttt{NSProblem}}.
Moreover, {\fontsize{11pt}{11pt}\texttt{TurbSolver}} needs an object in the {\fontsize{11pt}{11pt}\texttt{NSSolver}}
hierarchy to solve the NS equations during the iterations of the total
system of PDEs. As mentioned in Section~\ref{sec:NS:problem:solver},
the {\fontsize{11pt}{11pt}\texttt{self.nu}} variable in an NS solver must now point to the finite
element function representing $\nu + \nu_T$ in the RANS model.

We have two basic choices when implementing a RANS model, either to
develop a specific implementation tailored to a particular model, or
to make a general toolbox for a system of PDEs.  The former approach
is exemplified in the next section for a $k$--$\epsilon$ model,
while the latter is discussed in Section~\ref{sec:RANS:generalize}.
To implement a $k$--$\epsilon$ model, one naturally makes a subclass
{\fontsize{11pt}{11pt}\texttt{KEpsilon}} in the {\fontsize{11pt}{11pt}\texttt{TurbSolver}} hierarchy. Some tasks are specific
to the $k$--$\epsilon$ model in question and are better distributed to
subclasses like {\fontsize{11pt}{11pt}\texttt{StandardKE}} or {\fontsize{11pt}{11pt}\texttt{LowReynolds}}.
The choice between the three low-Reynolds
models is made in {\fontsize{11pt}{11pt}\texttt{LowReynolds}} whereas some of the data structures
are defined in subclasses for either a coupled or segregated approach. In
Figure~\ref{fig:Turbhier} we also sketch the possibility of having a
{\fontsize{11pt}{11pt}\texttt{TwoEquationModel}} class with functionality common to all two-equation
models.

A {\fontsize{11pt}{11pt}\verb!setup!} method defines the data structures and forms needed for
a solution of the $k$ and $\epsilon$ equations. The code is similar to
the {\fontsize{11pt}{11pt}\texttt{setup}} method for the NS solver classes.  Definition of the
specific forms is performed through the {\fontsize{11pt}{11pt}\texttt{Scheme}} class hierarchy. Any
solver class has a {\fontsize{11pt}{11pt}\texttt{schemes}} attribute which holds a dictionary of
all the needed schemes. A coupled low-Reynolds model will have a scheme
{\fontsize{11pt}{11pt}\texttt{'ke'}} for solving the $k$ and $\epsilon$ equations, and a segregated
model will have schemes {\fontsize{11pt}{11pt}\texttt{'k'}} and {\fontsize{11pt}{11pt}\texttt{'e'}} for the individual $k$
and $e$ equations. In addition, there is a list of schemes for all the
derived quantities, such as $\nu_T$, $f_\mu$, $f_2$, etc.

All scheme objects are declared through the method {\fontsize{11pt}{11pt}\verb!define!}, which
typically will be called as the final task of the {\fontsize{11pt}{11pt}\verb!setup!}
method. It is possible to change the composition of scheme
objects at run-time and simply rerun {\fontsize{11pt}{11pt}\texttt{define}}, without having
to reinitialize function spaces, test and trial functions,
and unknowns.
\subsubsection{Defining a specific two-equation model}
\label{sec:RANS:schemes}

The equations of all turbulence models are defined by subclasses of
{\fontsize{11pt}{11pt}\texttt{Scheme}} and can be transparently used with Picard or Newton
iterations, as previously exemplified for a coupled NS solver in
Section~\ref{sec:NS:iterschemes}.  Here is an outline of a coupled
$k$-$\epsilon$ model:
\begin{Verbatim}[fontsize=\fontsize{11pt}{11pt},tabsize=8,baselinestretch=1.0]
class KEpsilonCoupled(Scheme):
    def __init__(self, solver, unknown):
        Scheme.__init__(self, solver, ...)
        form_args = vars(solver).copy()
        if self.prm['iteration_type'] == 'Picard':
            F = self.form(**form_args)
            self.a, self.L = lhs(F), rhs(F)
        elif self.prm['iteration_type'] == 'Newton':
            form_args['k_'], form_args['e_'] = solver.k, solver.e
            F = self.form(**form_args)
            ke_, ke = unknown, solver.ke
            F_ = action(F, function=ke_)
            J_ = derivative(F_, ke_, ke)
            self.a, self.L = J_, -F_
\end{Verbatim}
\noindent
As in the {\fontsize{11pt}{11pt}\texttt{CoupledBase}} constructor for the NS schemes, we send all
attributes in the solver class as keyword arguments to the {\fontsize{11pt}{11pt}\texttt{form}}
methods. Most of these arguments are never used and are absorbed by a
final {\fontsize{11pt}{11pt}\texttt{**kwargs}} argument, but the number of variables needed to
define a form is still quite substantial:
\begin{Verbatim}[fontsize=\fontsize{11pt}{11pt},tabsize=8,baselinestretch=1.0]
class Steady_ke_1(KEpsilonCoupled):

    def form(self, k, e, v_k, v_e,  # Trial and TestFunctions
                   k_, e_, nut_, u_, Sij_, E0_, f2_, D_, # Functions/forms
                   nu, e_d, sigma_e, Ce1, Ce2, **kwargs): # Constants

        Fk = (nu + nut_)*inner(grad(v_k), grad(k))*dx \
            + inner(v_k, dot(grad(k), u_))*dx \
            - 2.*inner(grad(u_), Sij_)*nut_*v_k*dx \
            + (k_*e*e_d + k*e_*(1. - e_d))*(1./k_)*v_k*dx + v_k*D_*dx

        Fe = (nu + nut_*(1./sigma_e))*inner(grad(v_e), grad(e))*dx \
            + inner(v_e, dot(grad(e), u_))*dx \
            - (Ce1*2.*inner(grad(u_), Sij_)*nut_*e_ \
            - f2_*Ce2*e_*e)*(1./k_)*v_e*dx - E0_*v_e*dx

        return Fk + Fe
\end{Verbatim}
\noindent
Variants of this form, with different linearizations, are defined
similarly. By a proper construction of class names, based on
user-given parameters, the factory function for creating the right
scheme object can be coded in one line with {\fontsize{11pt}{11pt}\texttt{eval}}, as exemplified
in {\fontsize{11pt}{11pt}\texttt{NSCoupled.define}}.  On the contrary, registering a user-defined
scheme in a library coded in a statically typed language (Fortran, C,
C++, Java, or C\#) requires either an extension of the many {\fontsize{11pt}{11pt}\texttt{switch}}
or {\fontsize{11pt}{11pt}\texttt{if-else}} statements of a factory function in the library, or
sophisticated techniques to overcome the constraints of static typing.

The turbulence solver class mimics most of the code presented for the
{\fontsize{11pt}{11pt}\texttt{NSCoupled}} class. That is, we must define function spaces for $k$ and
$\epsilon$, and a compound (mixed) space for the coupled system.  The primary
unknown in this system, called {\fontsize{11pt}{11pt}\texttt{ke}}, and its {\fontsize{11pt}{11pt}\texttt{Function}}
counterpart {\fontsize{11pt}{11pt}\verb!ke_!}, are both defined similarly to {\fontsize{11pt}{11pt}\texttt{up}}
and {\fontsize{11pt}{11pt}\verb!up_!} in class {\fontsize{11pt}{11pt}\texttt{NSCoupled}}.  A considerable extension,
however, is the need to define all the parameters and quantities that
enter the turbulence model. For the {\fontsize{11pt}{11pt}\texttt{form}} method above to work,
these quantities must be available as attributes {\fontsize{11pt}{11pt}\verb!fmu_!}, {\fontsize{11pt}{11pt}\verb!f2_!},
etc., in the solver class so that the {\fontsize{11pt}{11pt}\verb!form_args!} dictionary contains
these names and can feed them to the {\fontsize{11pt}{11pt}\texttt{form}} method.  Details on the
definitions of turbulence quantities will appear later.

\subsubsection{General systems of turbulence PDEs}
\label{sec:RANS:generalize}

The briefly described classes for the $k$-$\epsilon$ model are very
similar to the corresponding classes for the NS schemes, and in fact to
all other turbulence models.  The only difference is the name of the
primary unknowns, their corresponding variable names in the solver class
and their coupling.  An obvious idea is to parameterize the names of the
primary unknowns in turbulence models and create code that is common.
This makes the code for adding a new model dramatically shorter.

\begin{figure}
  \center{\includegraphics[width=1.0\textwidth]{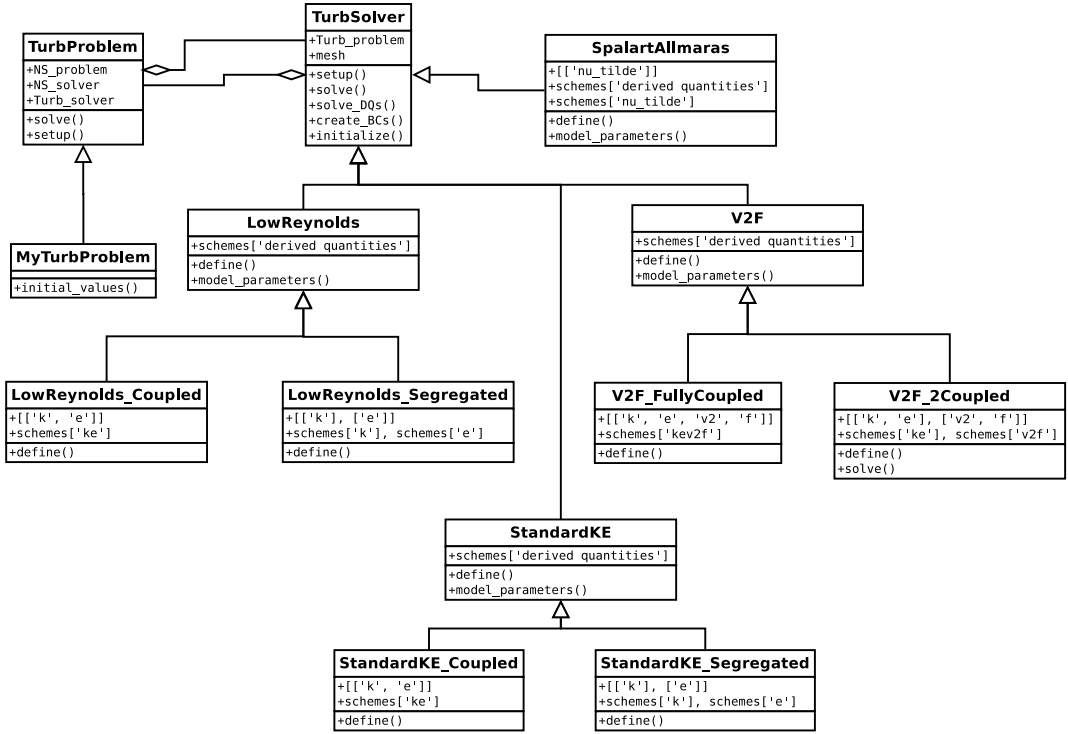}}
  \caption{Sketch of a modified and extended collection of turbulence
  problem and solver classes. For better illustration, the lowest level
  of solver classes show the value of the system composition attribute
  as lists.}
\label{fig:Turbhier2}
\end{figure}

For the solution of a general system of PDEs, we introduce a list, here
called
{\fontsize{11pt}{11pt}\verb!system_composition!}, containing the names of the primary unknowns
in the system and how they are grouped into subsystems that are to be
solved simultaneously.  For example, {\fontsize{11pt}{11pt}\texttt{[['k', 'e']]}} defines only
one subsystem consisting of the primary unknowns {\fontsize{11pt}{11pt}\texttt{k}} and {\fontsize{11pt}{11pt}\texttt{e}}
to be solved for in a coupled fashion (see {\fontsize{11pt}{11pt}\verb!LowReynolds_Coupled!}
in Figure~\ref{fig:Turbhier2}).  The list {\fontsize{11pt}{11pt}\texttt{[['k'], ['e']]}} defines
two subsystems, one for {\fontsize{11pt}{11pt}\texttt{k}} and one for {\fontsize{11pt}{11pt}\texttt{e}}, and is the relevant
specification for a segregated formulation of a $k$--$\epsilon$ model
(e.g., {\fontsize{11pt}{11pt}\verb!LowReynolds_Segregated!} in Figure~\ref{fig:Turbhier2}).
A fully coupled $v^2$-$f$ model \citep{book:durbin} is specified by
the list {\fontsize{11pt}{11pt}\verb![['k', 'e', 'v2', 'f']]!}, while the common segregated
strategy of solving a coupled $k$-$\epsilon$ system and a coupled
$v^2$-$f$ system is specified by {\fontsize{11pt}{11pt}\texttt{[['k', 'e'], ['v2', 'f']]}}. Using
the {\fontsize{11pt}{11pt}\verb!system_composition!} list and a few simple naming rules, we
will now illustrate how all tasks of creating relevant function spaces,
data structures, boundary conditions and even initialization, can be
fully automated in the superclass {\fontsize{11pt}{11pt}\texttt{TurbSolver}} for most systems.
Figure~\ref{fig:Turbhier2} shows the class hierarchy where the superclass
{\fontsize{11pt}{11pt}\texttt{TurbSolver}} performs most of the work and the individual models
need merely to implement bare necessities like {\fontsize{11pt}{11pt}\verb!model_parameters!}
and {\fontsize{11pt}{11pt}\texttt{define}} to set up the {\fontsize{11pt}{11pt}\texttt{schemes}} dictionary.

From the {\fontsize{11pt}{11pt}\verb!system_composition!} list we can define the name of a
subsystem as a simple concatenation of the unknowns in the subsystem,
e.g., {\fontsize{11pt}{11pt}\texttt{ke}} for a coupled $k$-$\epsilon$ model and {\fontsize{11pt}{11pt}\texttt{kev2f}} for
a fully coupled $v^2$-$f$ model. We need to create unknowns for these
concatenated names as well as {\fontsize{11pt}{11pt}\texttt{Functions}} for the names and all
individual unknowns. This can be done compactly as shown below.

We will now go into details of the abstract code required to automate common
tasks.  Dictionaries, indexed by the name of an unknown
(a single unknown such as {\fontsize{11pt}{11pt}\texttt{k}}, or a compound name for the unknown in
a subsystem such as {\fontsize{11pt}{11pt}\texttt{ke}}), are introduced: {\fontsize{11pt}{11pt}\texttt{V}} for the function spaces, {\fontsize{11pt}{11pt}\texttt{v}}
for the test functions, {\fontsize{11pt}{11pt}\texttt{q}} for the trial functions, {\fontsize{11pt}{11pt}\verb!q_!} for the
{\fontsize{11pt}{11pt}\texttt{Function}} objects holding the last computed approximation to {\fontsize{11pt}{11pt}\texttt{q}},
{\fontsize{11pt}{11pt}\verb!q_1!} for the {\fontsize{11pt}{11pt}\texttt{Function}} objects holding the unknowns at the
previous time step (and {\fontsize{11pt}{11pt}\verb!q_2!}, {\fontsize{11pt}{11pt}\verb!q_3!}, for older time steps,
if necessary), and {\fontsize{11pt}{11pt}\texttt{q0}} for the initial conditions.  For example,
{\fontsize{11pt}{11pt}\texttt{q['e']}} holds the trial function for $\epsilon$ in a $k$-$\epsilon$ model,
{\fontsize{11pt}{11pt}\texttt{v['e']}} is the corresponding test function, {\fontsize{11pt}{11pt}\texttt{V['e']}} is the
corresponding function space, {\fontsize{11pt}{11pt}\texttt{q['ke']}} holds the trial function
(in a space {\fontsize{11pt}{11pt}\texttt{V['ke']}}) for the compound unknown ($k$, $\epsilon$)
in a coupled $k$-$\epsilon$ formulation, {\fontsize{11pt}{11pt}\verb!q_['ke']!} (in a space
{\fontsize{11pt}{11pt}\texttt{V['ke']}}) holds the corresponding computed finite element function,
and so on.

The constructor takes the system composition and creates lists of all
names of all unknowns and the names of the subsystems:
\begin{Verbatim}[fontsize=\fontsize{11pt}{11pt},tabsize=8,baselinestretch=1.0]
class TurbSolver:
    def __init__(self, system_composition, problem, parameters):
        self.system_composition = system_composition
        self.system_names = [];  self.names = []

        for sub_system in self.system_composition:
            self.system_names.append(''.join(sub_system) )
            for name in sub_system:
                self.names.append(name)

\end{Verbatim}
\noindent
Defining the function spaces for each unknown and each compound unknown
in subsystems is done by a dict comprehension:
\begin{Verbatim}[fontsize=\fontsize{11pt}{11pt},tabsize=8,baselinestretch=1.0]
    def define_function_spaces(self):
        mesh = self.Turb_problem.NS_problem.mesh
        self.V = {name: FunctionSpace(mesh, 'Lagrange',
                        self.prm['degree'][name])
                  for name in self.names + ['dq']}

        for sub_sys, sys_name in \
            zip(self.system_composition, self.system_names):
            if len(sub_sys) > 1:  # more than one PDE in the system?
                self.V[sys_name] = MixedFunctionSpace(
                    [self.V[name] for name in sub_sys])

\end{Verbatim}
\noindent
For a coupled $k$-$\epsilon$ model, the first assignment to
{\fontsize{11pt}{11pt}\texttt{self.V}} creates the spaces {\fontsize{11pt}{11pt}\texttt{self.V['k']}}, {\fontsize{11pt}{11pt}\texttt{self.V['e']}}
and {\fontsize{11pt}{11pt}\texttt{self.V['dq']}}, while the next for loop creates the mixed space
{\fontsize{11pt}{11pt}\texttt{self.V['ke']}}. The {\fontsize{11pt}{11pt}\texttt{self.V['dq']}} object holds the space for
derived quantities and is always added to the collection of spaces.

The test, trial, and finite element functions for the compound unknowns
are readily constructed by:
\begin{Verbatim}[fontsize=\fontsize{11pt}{11pt},tabsize=8,baselinestretch=1.0]
    def setup_subsystems(self):
        V, sys_names, sys_comp = \
               self.V, self.system_names, self.system_composition

        q   = {name: TrialFunction(V[name]) for name in sys_names}
        v   = {name: TestFunction(V[name]) for name in sys_names}
        q_  = {name: Function(V[name]) for name in sys_names}
\end{Verbatim}
\noindent
The quantities corresponding to the individual unknowns
are obtained by splitting objects for compound unknowns. Typically,
\begin{Verbatim}[fontsize=\fontsize{11pt}{11pt},tabsize=8,baselinestretch=1.0]
        for sub_sys, sys_name in zip(sys_comp, sys_names):
            if len(sub_sys) > 1:  # more than one PDE in the system?
                q_.update({sub_sys[i]: f[i] \
                           for i, f in enumerate(q_.split())}
\end{Verbatim}
\noindent
with a similar splitting of {\fontsize{11pt}{11pt}\texttt{q}}, {\fontsize{11pt}{11pt}\texttt{v}}, etc.  Finally, these
dictionaries are stored as class attributes:
\begin{Verbatim}[fontsize=\fontsize{11pt}{11pt},tabsize=8,baselinestretch=1.0]
        self.v = v; self.q = q; self.q_ = q_
\end{Verbatim}
\noindent
It is also convenient to create solver attributes with the same names
as the keys in these dictionaries. That is, in a coupled $k$-$\epsilon$
model we make the short form {\fontsize{11pt}{11pt}\verb!self.k_!} for {\fontsize{11pt}{11pt}\verb!self.q_['k']!},
{\fontsize{11pt}{11pt}\verb!v_ke!} for {\fontsize{11pt}{11pt}\verb!self.v['ke']!}, and similarly:
\begin{Verbatim}[fontsize=\fontsize{11pt}{11pt},tabsize=8,baselinestretch=1.0]
        for key, value in v .items(): setattr(self, 'v_'+key, value)
        for key, value in q .items(): setattr(self, key, value)
        for key, value in q_.items(): setattr(self, key+'_', value)
\end{Verbatim}
\noindent
A dictionary {\fontsize{11pt}{11pt}\verb!self.x_!} for holding the unknown vectors in the various
linear systems are created in a similar way.  To summarize, the ideas
of the solver classes for NS problems and specific turbulence problems
are followed, but unknowns are parameterized by names in
dictionaries, with these names as keys, to hold the key objects.
Class attributes based on the names refer to the dictionary elements,
so that a solver class has attributes for trial and test functions,
finite element functions, etc., just as in the NS solver classes.
These class attributes are required when subclasses of {\fontsize{11pt}{11pt}\texttt{Scheme}} define
variational forms by sending solver attributes to a {\fontsize{11pt}{11pt}\texttt{scheme}} method
(see Sections~\ref{sec:NS:iterschemes} and \ref{sec:RANS:schemes}). For
example, when a {\fontsize{11pt}{11pt}\texttt{scheme}} method needs a parameter {\fontsize{11pt}{11pt}\verb!e_!} in the
form, the object {\fontsize{11pt}{11pt}\verb!solver.e_!} is sent as parameter ({\fontsize{11pt}{11pt}\texttt{solver}} being
the solver object), and this object is actually {\fontsize{11pt}{11pt}\verb!solver.q_['e']!}
as created in the code segments above, perhaps by splitting the compound
function {\fontsize{11pt}{11pt}\verb!solver.q_['ke']!} into its subfunctions.

With the names of the unknown parameterized, it becomes natural to also
create common code for the scheme classes associated with turbulence
models. We introduce a subclass {\fontsize{11pt}{11pt}\texttt{TurbModel}} of {\fontsize{11pt}{11pt}\texttt{Scheme}} that
carries out the tasks shown for the {\fontsize{11pt}{11pt}\texttt{KEpsilonCoupled}} class above,
but now for a general system of PDEs:
\begin{Verbatim}[fontsize=\fontsize{11pt}{11pt},tabsize=8,baselinestretch=1.0]
class TurbModel(Scheme):
    def __init__(self, solver, sub_system):
        sub_name = ''.join(sub_system)
        Scheme.__init__(self, solver, sub_system, ...)
        form_args = vars(solver).copy()

        if self.prm['iteration_type'] == 'Picard':
            F = self.scheme(**form_args)
            self.a, self.L = lhs(F), rhs(F)
        elif self.prm['iteration_type'] == 'Newton':
            for name in sub_system:
                # switch from Function to TrialFunction:
                form_args[name+'_'] = solver.q[name]
            F = self.scheme(**form_args)
            u_ = solver.q_[sub_name]
            F_ = action(F, function=u_)
            u = solver.q[sub_name]
            J_ = derivative(F_, u_, u)
            self.a, self.L = J_, -F_
\end{Verbatim}
\noindent
Note that {\fontsize{11pt}{11pt}\texttt{u}} denotes a general unknown (e.g., {\fontsize{11pt}{11pt}\texttt{k}}, {\fontsize{11pt}{11pt}\texttt{e}},
or~{\fontsize{11pt}{11pt}\texttt{ke}}) when automatically setting up the Newton system.

We now illustrate a specific subclass of a turbulence model.  Considering
a low-Reynolds model, we may create a subclass {\fontsize{11pt}{11pt}\texttt{LowReynolds}}
with a {\fontsize{11pt}{11pt}\texttt{define}} method that sets up a list of necessary derived
quantities to be computed in the NS solver ($\strainr$, $E$) and all
the derived quantities entering the low-Reynolds turbulence models
(cf.~Table~\ref{tab:lowre}).  These depend on the specific model,
whose name is available through the parameters dictionary in the
turbulence problem class. The coding of the {\fontsize{11pt}{11pt}\texttt{define}} method in class
{\fontsize{11pt}{11pt}\texttt{LowReynolds}} reads:
\begin{Verbatim}[fontsize=\fontsize{11pt}{11pt},tabsize=8,baselinestretch=1.0]
def define(self):
    V = self.V['dq'] # space for derived quantities
    DQ, DQ_NoBC = DerivedQuantity, DerivedQuantity_NoBC  # short forms
    NS = self.Turb_problem.NS_solver
    model = self.Turb_problem.prm['model']

    ns = dict(u_=NS.u_)
    NS.schemes['derived quantities'] = [
        DQ(NS, 'Sij_', NS.S, 'strain_rate(u_)', ns),
        ...]
    self.Sij_ = NS.Sij_; self.d2udy2_ = NS.d2udy2_
    ns = vars(self)   # No copy - derived quantities will be added
    self.schemes['derived quantities'] = dict(
      LaunderSharma=lambda :[
        DQ_NoBC(self, 'D_', V, 'nu/2./k_*inner(grad(k_), grad(k_))', ns),
        DQ(self, 'fmu_', V, 'exp(-3.4/(1. + (k_*k_/nu/e_)/50.)**2)', ns),
        DQ(self, 'f2_' , V, '1. - 0.3*exp(-(k_*k_/nu/e_)**2)', ns),
        DQ(self, 'nut_', V, 'Cmu*fmu_*k_*k_*(1./e_)', ns),
        DQ(self, 'E_'  , V, '2.*nu*nut_*dot(d2udy2_, d2udy2_)', ns)],
      JonesLaunder=lambda :[
        DQ_NoBC(self, 'D_', V, 'nu/2./k_*inner(grad(k_), grad(k_))', ns),
        DQ(self, 'fmu_', V, 'exp(-2.5/(1. + (k_*k_/nu/e_)/50.))', ns),
        DQ(self, 'f2_' , V, '(1. - 0.3*exp(-(k_*k_/nu/e_)**2))', ns),
        DQ(self, 'nut_', V, 'Cmu*fmu_*k_*k_*(1./e_)', ns),
        DQ(self, 'E_'  , V, '2.*nu*nut_*dot(d2udy2_, d2udy2_)', ns)],
      Chien=lambda :[...]
      )[model]()
    TurbSolver.define(self)
\end{Verbatim}
\noindent
A significant number of constants are involved in the expressions for
many of the derived quantities. These constants can be defined through
dictionaries in another method:
\begin{Verbatim}[fontsize=\fontsize{11pt}{11pt},tabsize=8,baselinestretch=1.0]
    def model_parameters(self):
        model = self.Turb_problem.prm['model']

        self.model_prm = dict(Cmu=0.09, sigma_e=1.30,
            sigma_k=1.0, e_nut=1.0, e_d = 0.5, f1 = 1.0)
        Ce1_Ce2 = dict(
            LaunderSharma=dict(Ce1=1.44, Ce2=1.92),
            JonesLaunder= dict(Ce1=1.55, Ce2=2.0),
            Chien=        dict(Ce1=1.35, Ce2=1.80))
        self.model_prm.update(Ce1_Ce2[model])
        # wrap in Constant objects:
        for name in self.model_prm:
            self.model_prm[name] = Constant(self.model_prm[name])
        # store model_prm objects as class attributes:
        self.__dict__.update(self.model_prm)
\end{Verbatim}
\noindent
Dictionaries are useful for storing the data, but in the scheme
objects defining the forms we need a constant like {\fontsize{11pt}{11pt}\texttt{Cmu}} as an
attribute in the class, and this is accomplished by simply updating the
{\fontsize{11pt}{11pt}\verb!__dict___!} dictionary. Also note that constants should be wrapped
in {\fontsize{11pt}{11pt}\texttt{Constant}} objects if they enter variational forms written in
UFL. That way constants may be changed without the need to recompile
UFL forms.

Subclasses of {\fontsize{11pt}{11pt}\texttt{LowReynolds}} define a segregated or a coupled scheme.
For example,
\begin{Verbatim}[fontsize=\fontsize{11pt}{11pt},tabsize=8,baselinestretch=1.0]
class LowReynolds_Coupled(LowReynolds):
    def __init__(self, problem, parameters):
        LowReynolds.__init__(self, system_composition=[['k','e']],
                             problem=problem, parameters=parameters)

    def define(self):
        LowReynolds.define(self)
        classname = self.prm['time_integration'] + '_ke_' + \
                    str(self.prm['scheme']['ke'])
        self.schemes['ke'] = eval(classname)\
                    (self, self.system_composition[0])
\end{Verbatim}
\noindent
Two actions are performed: definition of the subsystems to be solved
(here the $k$-$\epsilon$ system); and creation of the scheme class that
defines the variational form. If the user has set the solver parameters
\begin{Verbatim}[fontsize=\fontsize{11pt}{11pt},tabsize=8,baselinestretch=1.0]
turbsolvers.parameters['time_integration'] ='Steady'
turbsolvers.parameters['scheme']['ke'] = 1
\end{Verbatim}
\noindent
the {\fontsize{11pt}{11pt}\texttt{classname}} variable becomes {\fontsize{11pt}{11pt}\verb!Steady_ke_1!}, and the
corresponding class was shown in Section~\ref{sec:RANS:schemes} (except
that {\fontsize{11pt}{11pt}\verb!Steady_ke_1!} is now derived from {\fontsize{11pt}{11pt}\texttt{TurbModel}} and not the
specialized {\fontsize{11pt}{11pt}\texttt{KEpsilonCoupled}}).

A segregated solution approach to low-Reynolds models is implemented in a
subclass {\fontsize{11pt}{11pt}\verb!LowReynolds_Segregated!}, where the system composition reads
{\fontsize{11pt}{11pt}\texttt{[['k'],['e']]}},  {\fontsize{11pt}{11pt}\texttt{self.schemes['k']}} is set to some steady or
transient scheme object for the $k$ equation and {\fontsize{11pt}{11pt}\texttt{self.schemes['e']}}
is set to a similar object defining the form in the $\epsilon$ equation.

Several turbulence models have already been implemented in cbc.rans
using the generalized approach, and appear in Figure~\ref{fig:Turbhier2}:
the three low-Reynolds models as described, a standard $k$-$\epsilon$
model with wall functions, a Spalart-Allmaras one-equation model, a fully
coupled $v^2$-$f$ model and a $v^2$-$f$ model divided into a coupled
$k$-$\epsilon$ system and a coupled $v^2$-$f$ system.  For each of these
models, various {\fontsize{11pt}{11pt}\texttt{scheme}} methods in subclasses of {\fontsize{11pt}{11pt}\texttt{TurbModel}}
define various linearizations.

Another class of models involve equations for each of the six components
of the Reynolds stress tensor $\R$.  Minimal effort is required to make
the creation of function spaces in {\fontsize{11pt}{11pt}\texttt{TurbSolver}} work with vectors
or even second order tensors since UFL has support for both vector and
tensor-valued function spaces.  In fact, a coupled scheme for computing
all components of the Reynolds stress tensor can be expressed using
computer code which mirrors the mathematical tensor notation, just as
for the previously introduced scalar equations.

A particularly interesting application of the described framework is the
implementation of structure-based turbulence models \citep{Kassinos:00},
which introduce new turbulence measures and models, and lead to even
larger larger systems of PDEs than found in more common Reynolds stress
models. Such applications are in preparation by the authors.
\section{Numerical examples}
\label{RANS:results}

To demonstrate the framework, we consider two numerical examples.
The complete computer code for the presented examples can be found in
\citet{www:cbc.rans}.

\subsection{Fully developed turbulent channel}

Statistically one-dimensional, fully developed channel flow between
two parallel plates is used often in the development and verification
of CFD codes.  We will use this problem to investigate the impact
of different approximate linearizations.  The flow for the channel
problem is characterized by a Reynolds number based on the friction
velocity $u_{\tau}=(\sqrt{\nu \partial u_{x}/ \partial y})_{\rm wall}$,
where $u_x$ is the velocity tangential to the wall and $y$ is the wall
normal direction. The friction-based Reynolds number is defined as
$Re_{\tau}=u_{\tau}h/\nu$, where $h$ is half the channel height. Here
$Re_{\tau}=395$, $u_{\tau}=0.05$ and $h=1$ are used.  For this problem,
$\u = \mathbf{0}$, $k=0$ and $\epsilon =0$ on walls, and periodic
boundary conditions are applied at the inflow and outflow.  The laminar
flow profile is used for initial velocity field, and $k$ and $\epsilon$
are initially set to~$0.01$.

We set up {\fontsize{11pt}{11pt}\texttt{problem}} classes for the laminar and turbulent cases,
as outlined in Section~\ref{sec:RANS:problem}.  A mesh with 50 elements
in the wall normal direction for half the channel height and 10 elements
in the stream-wise direction is used.  Linear basis functions are used
for all fields and the flow is driven by a constant pressure gradient.
Due to the equal order or the velocity and pressure function spaces,
the stabilized form \eqref{eq:NS:varform:coupled_stabilized} of the
equations are used, with a constant $\tau=0.01$.  Picard iterations
are used with under-relaxation factors of 0.8 and 0.6 for the NS and
turbulence equations, respectively.  The {\fontsize{11pt}{11pt}\texttt{problem}} class designed
for the channel flow has to return initial values for $\u$, $k$ and
$\epsilon$. The {\fontsize{11pt}{11pt}\texttt{problem}} class also defines two {\fontsize{11pt}{11pt}\texttt{SubDomains}}
that are used to identify the wall and the mapping between the
periodic in- and outlets.

The first example concerns the linearization of the dissipation term
$\epsilon$ in the $k$ equation \refeq{k:eq:LS}.  Since $\epsilon$
appears in the $k$ equation, an explicit treatment of $\epsilon_{-}$
may appear natural.  However, an implicit treatment of $\epsilon$
will normally contribute to enhanced stability.  Moreover, we also
introduce a different implicit discretization through a weighting of
$\epsilon_{-}k/k_{-}$ and $\epsilon$ in \refeq{def:e_d}. This term
is often used by segregated solvers as it adds terms to the diagonal
entries of the coefficient matrix, which improves the condition number
to the benefit of Krylov solvers. Here we will use a direct solver.
The purpose is to investigate the impact of the weighting factor $e_{d}$
on the convergence behavior when solving the nonlinear equations.
We remark that the weighted form \refeq{def:e_d} (see \eqref{def:e_d})
is of relevance only for a coupled solution procedure since $e_d \neq 0$
implies that $\epsilon$ is unknown in the $k$ equation.

The simulation of the channel problem for all models for a range of $e_d$
values is scripted and run by:
\begin{Verbatim}[fontsize=\fontsize{11pt}{11pt},tabsize=8,baselinestretch=1.0]
# Get relevant solvers and problems
nssolver = nssolvers.NSCoupled
nschannel = turbproblems.NS_channel
turbchannel = turbproblems.channel
turbsolver = turbsolvers.LowReynolds_Coupled
# Redefine relevant parameters dictionaries
prm_ns = nsproblems.parameters
prm_turb = turbproblems.parameters
prm_nssol = nssolvers.parameters
prm_turbsol = turbsolvers.parameters
prm_ns.update(dict(Nx=10, Ny=50))
prm_nssol = recursive_update(prm_nssol,
         dict(degree=dict(velocity=1), omega=0.8))
prm_turbsol.update(omega=0.6)
prm_turb.update(Re_tau=395, utau=0.05)
# Loop over all models and e_d
for model in ['LaunderSharma','JonesLaunder', 'Chien']:
    for e_d in linspace(0,1,5):
        prm_turb.update(Model=model)
        prm_ns.update(prm_turb)
        NS_problem = nschannel(prm_ns)
        NS_solver = nssolver(NS_problem, prm_nssol)
        NS_solver.setup()
        Problem = turbchannel(NS_problem, prm_turb)
        Solver = turbsolver(Problem, prm_turbsol)
        Solver.setup()
        Solver.e_d = Constant(e_d)
        Solver.define()
        Problem.solve(max_iter=50, max_err=1e-12)
        ... # Store results etc
\end{Verbatim}
\noindent

\begin{figure}
  \center
  \includegraphics[width=0.7\textwidth]{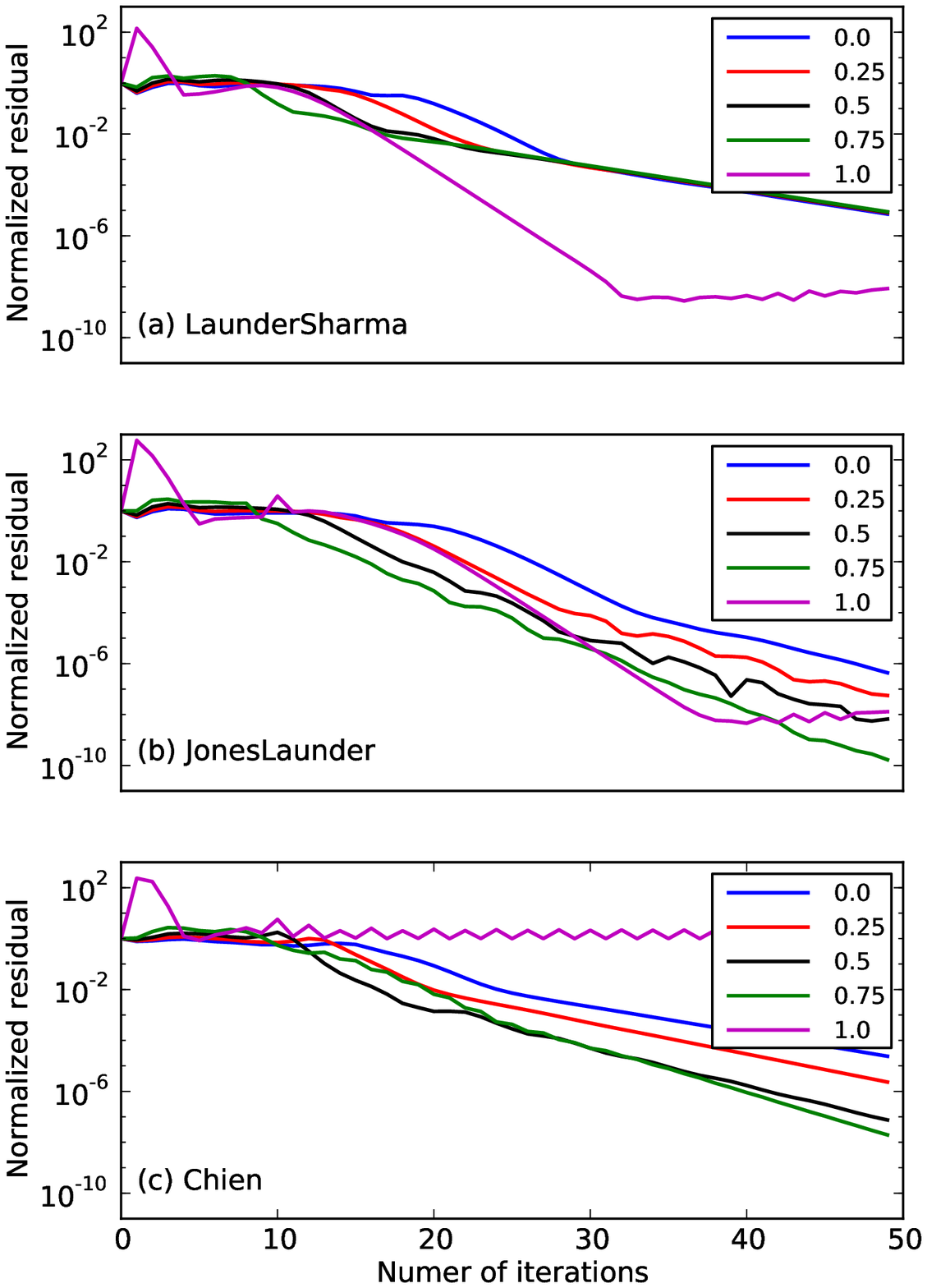}
  \caption{Plots of convergence history for a turbulent channel
  flow computed with {\fontsize{11pt}{11pt}\texttt{NSCoupled}}. The different lines represent
  the choice of weight $e_d$ as indicated in the legend. The norms of
  the $\u$ residuals (vertical axes) are plotted against the number of
  iterations (horizontal axes).  In (a)-(c) we plot the results of using
  the {\fontsize{11pt}{11pt}\texttt{LowReynolds}}$_{-}${\fontsize{11pt}{11pt}\texttt{Coupled}} solver with the LaunderSharma,
  JonesLaunder and Chien models, respectively. }
  \label{fig:implicit_dissipation}
\end{figure}

Figure~\ref{fig:implicit_dissipation} shows the convergence response
for the three low Reynolds number turbulence models discussed in
Section~\ref{math:models}. The fully coupled form using $e_d = 1$ is the
most efficient method for LaunderSharma and JonesLaunder, whereas it is
unstable for Chiens model. The ``segregated'' form $\epsilon_{-}k/k_{-}$
($e_d = 0$) is least efficient, whereas a blend of both forms seems
to be optimal as a default option for all models.  It is interesting
to note that even with a very poor (constant) initial guess for $k$
and $\epsilon$, and using rather robust under-relaxation factors, we
can generally find the solution in less than 50 iterations.
\subsection{Channel flow with and adverse pressure gradient}

We now consider a channel that has a bump on the lower wall,
thereby inducing an adverse pressure gradient that leads to
separation on the decelerating side. The bump geometry is described by
\citet{marquillie08}, who performed Direct Numerical Simulations of the
flow. The bump has also been studied experimentally at higher Reynolds
numbers~\citep{Kostas05,Bernard03}.

Adverse pressure gradient flows are notoriously difficult to model
with simple RANS models. One particular reason for this is that regular
wall function approaches do not work well, since the velocity profile
near the wall will be far from the idealized log-law (see Figure~4.4
in \citet{book:durbin}). Here we will employ the original four-equation
V2F model of \citet{Durbin:91} (see Fig.~\ref{fig:Turbhier2}) that does
not employ wall functions, but involves special boundary condition-like terms.
For numerical stability, these boundary conditions must be applied in a
coupled and implicit manner, which is hard (if not impossible) to do with
most commercial CFD software. We solve for the system composition
{\fontsize{11pt}{11pt}\texttt{[['k', 'e'], ['v2', 'f']]}}, where the scalar $v^2$ resembles a wall normal
stress and $f$ is a pressure redistribution parameter. The NS solver is
the same as in the previous example.

For this example, homogeneous Dirichlet boundary conditions are applied
to $\u$, $k$ and $v^2$ on walls.
Prescribed profiles
for $k, \epsilon, v^2$ and $f$ on the inlet, obtained from an independent
solution of a plain channel are used. The outlet
uses a pseudo-traction-free condition ($\nabla \u \cdot n - pn = 0$)
and zero normal derivative for all other turbulence quantities.
The ``boundary conditions'' used for $\epsilon$ and $f$ on walls are
\begin{equation}
 \epsilon = \frac{2 \nu k}{y^2} \text{ and } f = \frac{20 \nu^2 v^2}{y^4 \epsilon},
 \label{eq:e_wall}
\end{equation}
respectively, where $y$ is the wall normal distance. This ``boundary
condition'' cannot be set directly on the wall, because $y$ will be
zero. However, $k/y$ approaches the wall as $O(1)$, and can be evaluated
on the degrees of freedom close to the walls.  In practice we set this
boundary condition implicitly for $\epsilon$ and $f$ by manipulating the
rows of the coefficient matrix that represent the degrees of freedom of
elements in contact with a wall. The ``boundary value'' is computed
for the degrees of freedom furthest from the wall for
elements that are in contact with the wall. Within these elements,
the fields are kept constant. The distance to the nearest wall $y$
is computed through a stabilized Eikonal equation, implemented as
\begin{Verbatim}[fontsize=\fontsize{11pt}{11pt},tabsize=8,baselinestretch=1.0]
    F = sqrt(inner(grad(y), grad(y)))*v*dx - f*v*dx \
           + eps*inner(grad(y), grad(v))*dx,
\end{Verbatim}
\noindent
where {\fontsize{11pt}{11pt}\verb!f=Constant(1)!} and {\fontsize{11pt}{11pt}\verb!eps=Constant(0.01)!}. The same
{\fontsize{11pt}{11pt}\texttt{FunctionSpace}} as for the turbulence parameters is used and the only
assigned boundary conditions for $y$ are homogeneous Dirichlet on walls.
The stabilized Eikonal equation is solved in just a few iterations using
a Newton method.

\begin{figure}
\centering
\subfigure[]
{
\includegraphics[width=0.8\textwidth]{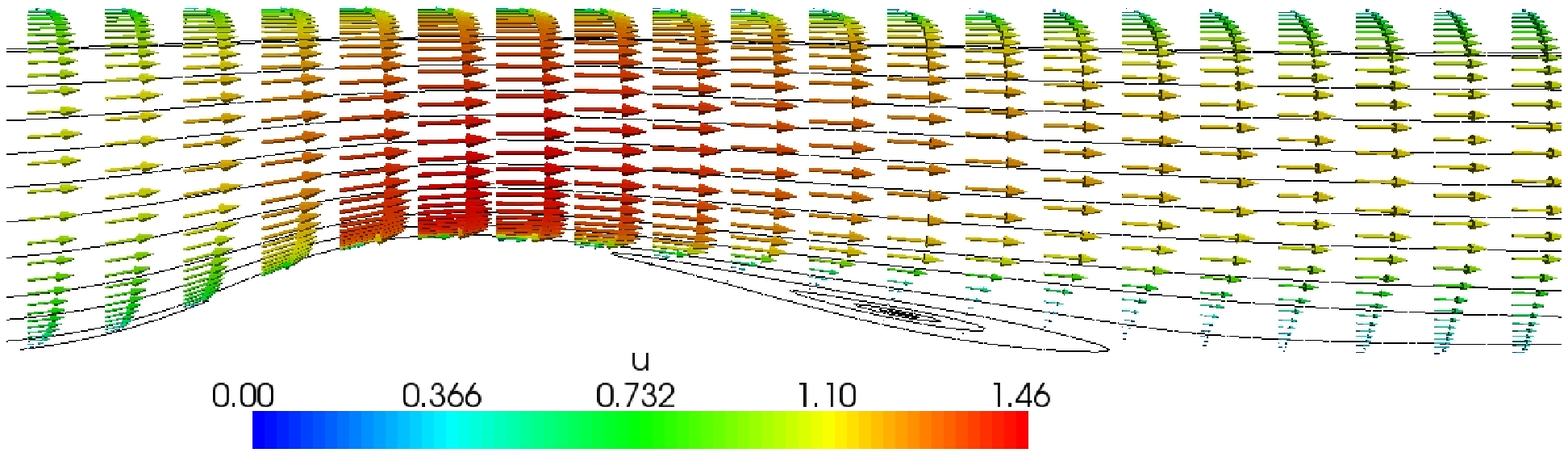}
\label{fig:psi_apbl}
}
\subfigure[]
{
\includegraphics[width=0.8\textwidth]{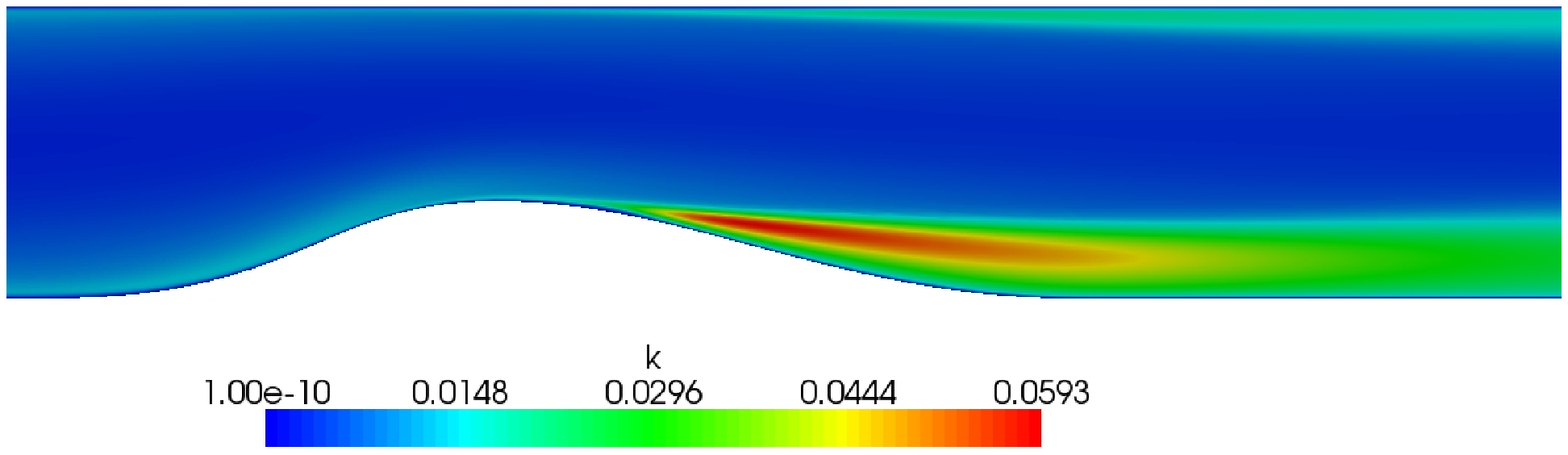}
\label{fig:k_apbl}
}
\subfigure[]
{
\includegraphics[width=0.8\textwidth]{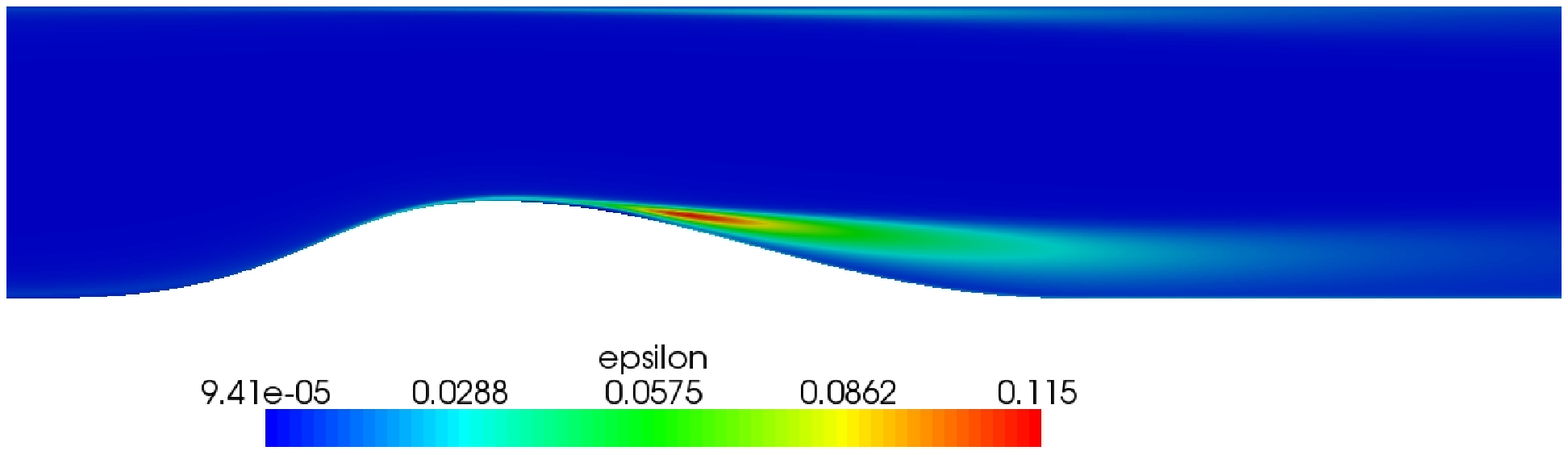}
\label{fig:e_apbl}
}
\caption{Computed quantities for flow past a bump using the $v^{2}-f$
model: (a) velocity vectors and streamfunction; (b) turbulent kinetic
energy contours [m$^2$/s$^{2}$]; and (c) rate of energy dissipation
[m$^2$/s$^3$].}
\label{fig:apbl}
\end{figure}

We create a solver similar to {\fontsize{11pt}{11pt}\verb!solve_nonlinear!} where it is iterated
between solving the NS-equations and the turbulence equations. Derived
quantities are also updated between each subsystem's solve.
The flow field is initialized using the channel solution and the flow
converges in about 50 iterations. Figure~\ref{fig:apbl} shows the
velocity and streamfunction (a), turbulent kinetic energy (b), and
rate of energy dissipation (c) in the near vicinity of the bump. The
streamfunction reveals that the flow separates on the decelerating
side of the bump, in good agreement with direct numerical simulations
of the same flow~\citep{marquillie08}.  The streamfunction shown in
Figure~\ref{fig:apbl}(a) has been computed in FEniCS through solving
the Poisson equation $\nabla^2 \psi = -\nabla \times \u$, which is
implemented as
\begin{Verbatim}[fontsize=\fontsize{11pt}{11pt},tabsize=8,baselinestretch=1.0]
    q   = TestFunction(V)
    psi = TrialFunction(V)
    n = FacetNormal(mesh)
    F = inner(grad(q), grad(psi))*dx \
        - inner(q, (u[1].dx(0) - u[0].dx(1)))*dx \
        + q*(n[1]*u[0] - n[0]*u[1])*ds.
\end{Verbatim}
\noindent
Here the boundary condition on the derivatives of $\psi$ (using
$ \nabla \times \psi = \u$) is set weakly
and it is noteworthy that this simple implementation applies for any
2 dimensional problem with any boundary.
This would also be difficult to do in most CFD software.
\section{Comments on computational efficiency}
\label{sec:comp:efficiency}

A key result from Section~\ref{sec:RANS:framework} is the convenient
specification of the equations which define numerical schemes and
how this specialized code is coupled with general code for handling an
arbitrary system of PDEs.  The user can freely choose between segregated
and coupled formulations, as well as Picard-type or Newton iterations.
A natural question is how this convenience and flexibility affects the
computational efficiency.

Scripted programming languages inevitably involve an overhead cost that
is greater than that for compiled languages.  To reconcile this, FEniCS
programs generate C++ code such that the solver carries out almost all
computations in C++ (although this is transparent to the user, as outlined
in Section~\ref{FEniCS:inside}).  Furthermore, representations of finite
element matrices and vectors can generated by a form compiler that are not
tractable by hand~\citep{KirbyLogg2006a}, and a range of domain-specific
automated optimizations are applied to reduce the number of floating point
operations~\citep{KirbyKnepleyEtAl2005a,oelgaard:2010}.  Therefore, we
expect our FEniCS-based RANS solvers to have a computational efficiency
superior to traditional, hand-written finite element solvers in~C++.
Comparisons with a state-of-the-art unstructured finite volume code
CDP from Stanford for a test case show that an optimized version of
our FEniCS-based NS solver is only approximately two times slower for
some laminar flow cases. This is despite scope remaining for further
optimizations of the FEniCS libraries and the extra flexibility offered
by the finite element framework.

The Python overhead of turbulence solvers is mostly due to callbacks
to Python from C++ for defining boundaries and initial conditions.
The latter are computed only once per simulation, while boundary
information is needed every time a linear system is computed. FEniCS has
constructs for avoiding callbacks to Python by using just-in-time compilation,
thereby eliminating the potential Python overhead.
\section{Concluding remarks}
\label{sec:conclusion}

We have presented a novel software framework for RANS modeling of
turbulent flow.  The framework targets researchers in computational
turbulence who want full flexibility with respect to composing PDE models,
coupling equations, linearizing equations and constructing iterative
strategies for solving the nonlinear equations arising in RANS models.

The use of Python and FEniCS to realize the framework yields compact,
high-level code, which in the authors' opinions provides greater
readability, flexibility and simplicity than what can be achieved
with C++ or Fortran 2003.  Throughout the text we have commented
upon object-oriented designs via class hierarchies versus generative
programming via stand-alone functions.  It is the authors' experiences
that Python leads to a design more biased toward the generative style than
does C++, perhaps because the generative style becomes so obvious in a
language with dynamic typing.  Although subclassing is a natural choice
for users to provide new implementations, we have emphasized stand-alone
functions as simpler and more flexible.  Classes and dictionaries are
used extensively in the code, but a disadvantage with Python, at least
when implementing mathematical formulae, is the {\fontsize{11pt}{11pt}\texttt{self}} prefix and
other disturbing syntax. This disadvantage is overcome partly by a clever
use of namespaces, as well as through the semi-automatic introduction
of local variables.

Our computational examples illustrate two tasks that are easy to
perform in the suggested framework, but usually hard to accomplish in
other types of CFD software, namely an investigation of various
linearizations in a family of turbulence models, and implementation of
the promising $v^2$-$f$ model with different degree of implicitness.

Many more features than shown in this work can be added to the
framework. For example, in FEniCS discontinuous Galerkin methods pose no
difficulty for the programmer.  FEniCS support for error control and
adaptivity is also a promising topic to include and explore, especially
since optimal mesh design is a major challenge in CFD.  Unsteady RANS
models with time dependency constitute an obvious extension, which
essentially consists in adding suitable finite difference schemes for the
time derivatives and a time loop in the code.  The flexibility offered
by FEniCS and the design of the RANS solvers makes the addition of
new functionality straightforward.

We believe that our combination of mathematical formulations and specific
code for the complicated class of PDE models addressed in this
work demonstrate the power of Python and FEniCS as an expressive
and human/computer-efficient software framework. Readers may use
our implementation ideas in a wide range of science and engineering
disciplines.
\bibliographystyle{model1b-num-names}
\bibliography{nsrefs}
\end{document}